\documentclass[12pt]{article}
\usepackage{amsmath}
\usepackage[dvips]{epsfig}
\usepackage{amsmath, amsthm, amssymb,bm}
\usepackage{xcolor}
\usepackage{multirow}
\usepackage{graphicx}
\usepackage{subfigure}
\usepackage{epstopdf}
\usepackage[round,sort&compress]{natbib}
\usepackage{caption}
\usepackage{listings}
\usepackage{makecell}
\usepackage{appendix}

\newcommand{\ignore}[1]{}{}
\newtheorem{lemma}{Lemma}

\newtheorem{theorem}{Theorem}
\newtheorem{corollary}{Corollary}

\numberwithin{equation}{section}
\theoremstyle{plain}

\makeatletter
\begin{document}
%-----------------------------------

\title{Diagonal Likelihood Ratio Test for  Equality of Mean Vectors in High-Dimensional Data}

\author{Zongliang Hu$^1$, Tiejun Tong$^{2,*}$ and Marc G. Genton$^{3}$ \\
\\
\small $^1$ College of Mathematics and Statistics, Shenzhen University, Shenzhen 518060, China \\
\small $^2$ Department of Mathematics, Hong Kong Baptist University, Hong Kong \\
\small $^3$ Statistics Program, King Abdullah University of Science and Technology, \\
{\small Thuwal 23955-6900, Saudi Arabia} \\
{\small $^*$Email:~tongt@hkbu.edu.hk}
}

\date{}

\maketitle

\begin{abstract}
\noindent
We propose a likelihood ratio test framework for testing normal mean vectors in high-dimensional data under two common  scenarios: the one-sample test and the two-sample test with equal covariance matrices. We derive the test statistics under the assumption that the covariance matrices follow a diagonal matrix structure. In comparison with the diagonal Hotelling's tests, our proposed test statistics display some interesting characteristics. In particular, they are a summation of the log-transformed squared  $t$-statistics rather than a direct summation of those components.  More importantly, to derive the asymptotic normality of our test statistics under the null and local alternative hypotheses, we do not need the requirement that the covariance matrices follow a diagonal matrix structure. As a consequence, our proposed test methods are very flexible and  readily applicable in
practice. Simulation studies and a real data analysis are also carried out to demonstrate the advantages of our likelihood ratio test methods.


\end{abstract}

\section{Introduction}\label{sec1}

In high-dimensional data analysis, it is often necessary to test whether a mean vector is equal to another vector in the one-sample case, or to test whether two mean vectors are equal to each other in the two-sample case. One such example is to test
whether two gene sets, or pathways, have equal expression levels under two different experimental conditions. Given the two normal random samples, $\bm{X}_{1}, \ldots, \bm{X}_{n_1} \in \mathbb{R}^p$ and
$\bm{Y}_{1}, \ldots, \bm{Y}_{n_2} \in \mathbb{R}^p$, one well-known method for testing whether their mean vectors are equal is Hotelling's $T^2$ test,
\begin{align}\label{HT-square}
T^2=\frac{n_1n_2}{n_1+n_2}(\bm{\bar{X}}-\bm{\bar{Y}})^T S^{-1}(\bm{\bar{X}}-\bm{\bar{Y}}),
\end{align}
where $\bm{\bar{X}}$ and $\bm{\bar{Y}}$ are the sample mean vectors and  $S$ is the pooled sample covariance matrix.
Hotelling's $T^2$ test is well-behaved and has been extensively studied in the classical low-dimensional setting.
However, this classic test may not perform well or may not even be applicable to high-dimensional data with a small sample size. Specifically, it suffers from the singularity problem because the sample covariance matrix $S$ is singular when the dimension is larger than the sample size.

To overcome the singularity problem in Hotelling's $T^2$ test, \citet{bai1996} replaced  the sample covariance matrix in (\ref{HT-square}) with the identity matrix, so that their test statistic is essentially the same as $(\bm{\bar{X}}-\bm{\bar{Y}})^T(\bm{\bar{X}}-\bm{\bar{Y}})$.
Following  their method, \citet{chen2010} and \citet{ahmad2014}
 proposed some $U$-statistics for testing whether two mean vectors are equal.
 These test methods were referred to as the {\it unscaled Hotelling's tests} in \citet{Dong2015}.
 As an alternative,
\citet{chen2011} and \citet{li2016} proposed replacing the inverse sample covariance matrix $S^{-1}$ with  the regularized estimator  $(S+\lambda I_p)^{-1}$ in Hotelling's test statistic, where $I_p$ is the identity matrix and $\lambda>0$ is a regularization parameter.
\citet{lopes2011} proposed a random projection technique to estimate the sample covariance matrix.
 Specifically, they replaced $S^{-1}$ in Hotelling's test statistic with ${E}_{R}^{-1}\{R(R^TSR)^{-1}R^T\}$, where $R$ is a random matrix of size $p\times k$ and ${E}_R(\cdot)$ is the expectation operator over the distribution.
The random projection technique was further explored by, for example, \citet{thulin2014}, \citet{srivastava2015}
and   \citet{wei2016}.  \citet{Dong2015} referred to the test methods in this category as the {\it regularized Hotelling's tests}.

In addition to the aforementioned methods, replacing the sample covariance matrix with a diagonal sample covariance matrix is another popular approach to improving Hotelling's $T^2$ test. In particular, \citet{Wu2006},
\citet{srivastava2008} and \citet{Srivastava2009} considered the {\it diagonal Hotelling's test statistic}:
\begin{align}\label{diag.stat}
T_{{\rm diag},2}^2=\frac{n_1n_2}{n_1+n_2}(\bm{\bar{X}}-\bm{\bar{Y}})^T \{{\rm diag}(S)\}^{-1}(\bm{\bar{X}}-\bm{\bar{Y}}).
\end{align}
Recently, \citet{srivastava2013},  \citet{feng2015}  and \citet{Carroll2015}  also considered the diagonal Hotelling's tests under the assumption of unequal covariance matrices.
Their test statistics essentially follow  the diagonal structure in (\ref{diag.stat}),
$(\bm{\bar{X}}-\bm{\bar{Y}})^T \{{\rm diag}(S_1)/n_1+{\rm diag}(S_2)/n_2\}^{-1}(\bm{\bar{X}}-\bm{\bar{Y}})$,  where $S_1$ and $S_2$ are  two sample covariance matrices.
\citet{park2013test} modified the diagonal Hotelling's test statistic (2) based on the idea of leave-out cross validation.
\citet{Dong2015} proposed a shrinkage-based Hotelling's test that replaced the diagonal elements of the  sample covariance matrix in (\ref{diag.stat}) with some improved variance estimates.
To summarize, the diagonal Hotelling's tests are popular in practice for several reasons. First, since a diagonal matrix is always invertible for nonzero variance estimates, the singularity problem
in the classic test is circumvented.
Second, the diagonal Hotelling's tests are scale transformation invariant tests.
As suggested by  \citet{park2013test}, the scale transformation invariant tests usually
provide a better performance than the orthogonal transformation invariant tests including the unscaled Hotelling's tests and the regularized Hotelling's tests, especially when the variances of the signal components are small and the variances of the noise components are large. Last but not least, a diagonal covariance matrix assumption is also popular  in the high-dimensional literature, e.g., in \citet{dudoit2002},
\citet{bickel2004} and \citet{huang2010}.

Note that Hotelling's test statistic originated  from the likelihood ratio test in the classical setting when $p$ is smaller than $n$. Recently, researchers have also applied the likelihood ratio
test method to analyze high-dimensional data.
For instance, \citet{Jiang2013} and \citet{Jiang2015} tested mean vectors and covariance matrices of normal distributions using the likelihood ratio test method,  under the setting that $p$ is smaller than $n$ but in a way that allows $p/n \to 1$. \citet{zhao2016} proposed a generalized high-dimensional likelihood ratio test for the normal mean vector by  a modified union-intersection method.
\citet{stadler2016} provided a high-dimensional
likelihood ratio test for the two-sample test based on sample splitting.

Following the diagonal matrix structure  and the likelihood ratio test method,  we propose a new test framework for high-dimensional data with a small sample size. Unlike the existing diagonal Hotelling's tests, in which the sample covariance matrix $S$ was directly replaced with the diagonal matrix ${\rm diag}(S)$, our likelihood ratio test statistics are a summation of the log-transformed squared $t$-statistics, rather than a direct summation of those components. When the sample size is small, the standard $t$ tests may be unreliable due to the unstable variance estimates. As a remedy, our proposed tests use the log-transformed squared $t$-statistics and, consequently, provide more stable test statistics so that type I error rates are better controlled for small sample sizes. We demonstrate by simulation that our proposed tests are robust in terms of controlling  the type I error rate at the nominal level in a wide range of settings.

The rest of the paper is organized as follows. In Section 2, we propose  the diagonal likelihood ratio test method  for the one-sample case. The asymptotic distributions of the test statistics are also derived as $p$ tends to infinity under the null and local alternative hypotheses, respectively. In Section 3, we propose the diagonal likelihood ratio test method  for the two-sample case and derive some asymptotic results, including the asymptotic null distribution and power.
In Section 4, we conduct simulation studies to evaluate the proposed tests and to compare them with existing methods. We apply the proposed tests to a real data example in Section 5, and  conclude the paper by providing a short summary and some future research directions in Section 6. The technical proofs are provided in the Appendices.

\section{One-Sample Test}\label{one-sample}
\subsection{Diagonal LRT statistic}
To illustrate the main idea of the diagonal likelihood ratio test method, we consider the one-sample test for a mean vector.
Let $\bm{X}_i=(X_{i1},X_{i2,},\ldots,X_{ip})^T, i=1,\ldots,n$, be independent and identically distributed (i.i.d.) random vectors from the multivariate normal distribution $N_{p}(\bm{\mu},\Sigma)$,
where $\bm{\mu}$ is the population mean vector and $\Sigma$ is the population covariance matrix. In the one-sample case, for a given vector $\bm{\mu}_0$, we test the hypothesis,
\begin{equation}\label{test1-1}
{H}_0: \bm{\mu}=\bm{\mu}_0 \text{~~versus~~} H_1: \bm{\mu}\neq \bm{\mu}_0.
\end{equation}

Our new likelihood ratio test statistic is based on the assumption that the covariance matrix follows a diagonal matrix structure, i.e., $\Sigma={\text{diag}}(\sigma_{11}^2,\ldots,\sigma_{pp}^2)$.
%In Appendix \ref{stat1},
In Appendix A.1, we show that the likelihood ratio test statistic for hypothesis (\ref{test1-1}) is
\begin{equation*}
\Lambda_n=\frac{\underset{\Sigma}{\mathrm{max}} L(\bm{\mu}_0,\Sigma)}
{\underset{\bm{\mu},\Sigma}{\mathrm{max}} L(\bm{\mu},\Sigma)}
=\frac{\prod_{j=1}^{p}\big\{\sum_{i=1}^{n}(X_{ij}-\bar{X}_j)^2 \big\}^{n/2}}
{\prod_{j=1}^{p}\big\{\sum_{i=1}^{n}(X_{ij}-\mu_{0j})^2\big\}^{n/2}},
\end{equation*}
where $\bar{X}_j=\sum_{i=1}^{n}{{X}_{ij}}/n$ are the sample means, and ${s}_j^2=\sum_{i=1}^{n}{({X}_{ij}-\bar{X}_j)^2}/(n-1)$ are the sample variances.
Taking the log transformation, we derive that
\begin{equation*}
-2 \mathrm{log}(\Lambda_n)=n\sum_{j=1}^{p}\mathrm{log} \bigl[1+{n}(\bar{X}_j-\mu_{0j})^2/\{(n-1)s_j^2 \}\bigr].
\end{equation*}
This suggests that the new test statistic is
\begin{align}\label{new.test1}
T_1 &=n\sum_{j=1}^{p}\log
\Bigl\{1+\frac{n(\bar{X}_j-\mu_{0j})^2}{(n-1)s_j^2}\Bigr\}
=n\sum_{j=1}^{p}\log
\Bigl(1+\frac{t_{nj}^2}{\nu_1} \Bigr),
\end{align}
where $t_{nj}=\sqrt{n}(\bar{X}_j-{\mu}_{0j})/s_{j}$
are the standard  $t$-statistics for the one-sample test
with $\nu_1=n-1$ degrees of freedom.  We refer to the diagonal likelihood ratio test statistic in (\ref{new.test1}) as the DLRT statistic.

 Under the null hypothesis, it is easy to verify that
$n \log\bigl(1+{t_{nj}^2}/{\nu_1} \bigr)=t_{nj}^2+O_p(1/n)$.
Further, we have $T_1=\sum_{j=1}^{p}t_{nj}^2 +O_p(p/n)$.
So if  $p$ increases at such a rate that  $p=o(n)$, then we have the following approximation:
\begin{equation*}
T_1 \approx
\sum_{j=1}^{p}t_{nj}^2
=n(\bm{\bar{X}}-\bm{\mu}_0)^T \{\mathrm{diag}(s_{1}^2,\ldots,s_{p}^2)\}^{-1}
(\bm{\bar{X}}-\bm{\mu}_0).
\end{equation*}
Thus, as a special case, our proposed DLRT statistic reduces to the diagonal Hotelling's test statistic in the one-sample case to which a direct summation of the squared $t$-statistics is applied.

\subsection{Null distribution}\label{Normal.1}
For ease of notation, let $U_{nj}=n\log( 1+t_{nj}^2/\nu_1)
$ for $j=1,\ldots,p$.
In this section, we derive the asymptotic null distribution of the proposed DLRT statistic. To derive the limiting distribution, we first present a lemma; the proof is in Appendix A.2.

\vskip 12pt
\begin{lemma}\label{lamm1}
For the  gamma function $\Gamma(x)=\int_{0}^{\infty}t^{x-1}e^{-t}dt$,
let $\Psi(x)= \Gamma '(x)/ \Gamma(x)$ be the digamma function.
Also, let $D(x)=\Psi\{(x+1)/2\}-\Psi(x/2)$,
$m_1=nD(\nu_1)$, and $m_2=n^2\{D^2(\nu_1)-2D'(\nu_1)\}$.
\begin{enumerate}
  \item[$(a)$] For any $n\geq2$, we have $E(U_{nj})= m_1$ and $\mathrm{Var}(U_{nj}) = m_2-m_1^2$.
  \item[$(b)$] As $n \to \infty$, we have  $E(U_{nj})\to 1$ and $\mathrm{Var}(U_{nj}) \to 2$.
\end{enumerate}
\end{lemma}

Despite  $T_1$ has an additive form of log-transformed squared $t$-statistics,
our derivation of its limiting distribution needs to account for the dependence among $\{U_{n1},\ldots,U_{np}\}$.
For example, the scaling parameter of $T_1$
may need to incorporate the information of ${\rm Cov}(U_{n,j},U_{n,j+k})$.
We therefore need additional assumptions when establishing the asymptotic normality of the DLRT statistic.
Let $\alpha(\mathcal{F},\mathcal{G})=
\sup\{ \|P(A \cap B)-P(A)P(B)\|{:}~
A \in \mathcal{F} , B\in \mathcal{G}\}$
be the strong mixing coefficient between
two $\sigma$ fields, $\mathcal{F}$ and $\mathcal{G}$, that measures the degree of dependence between the two $\sigma$ fields.
We also assume that the following two regularity conditions hold for the sequence $\{U_{nj}, j=1,2,\ldots\}$:
\begin{description}
\item[(C1)]  Let $\alpha(r)=\sup\{
    \alpha(\mathcal{F}_1^k,\mathcal{F}_{k+r}^p){:}~
1\leq k \leq p-r\}$, where $\mathcal{F}_{a}^b=\mathcal{F}_{a,n}^b=
\sigma\{U_{nj}: a\leq j \leq b\}$.
Assume that the  stationary sequence $\{ U_{nj} \}$  satisfies the strong mixing condition such that
${\alpha(r)\downarrow0}$ as
${r\rightarrow\infty}$, {where $\downarrow$ denotes} the monotone decreasing convergence.

\item[(C2)]
  Suppose that $\sum_{r=1}^{\infty}\alpha(r)^{\delta/(2+\delta)}< \infty$ for some $\delta>0$, and for any $k \geq 0$,

$\lim_{p \rightarrow \infty}\sum_{j=1}^{p-k}{\rm Cov}(U_{nj},U_{n,j+k})/{(p-k)}=\gamma(k)$ exists.
\end{description}
\vskip 5pt

The following theorem establishes the asymptotic distribution of the DLRT statistic under the null hypothesis.

\vskip 12pt
\begin{theorem}\label{Th1}
Let $\bm{X}_{1},\ldots,\bm{X}_{n}$ be i.i.d. random vectors from $N_{p}(\bm{\mu},\Sigma)$.
If the sequence $\{ U_{nj} \}$ is stationary and satisfies conditions (C1) and (C2), then under the null hypothesis,
we have for any fixed $n\geq 2$,
\begin{align*}
\frac{{T}_1-p m_1}{\tau_{1}\sqrt{p}}
~{\overset{\mathcal{D}}\longrightarrow}~N(0,1)
{\rm~~ as~~} p \rightarrow \infty
\end{align*}
 where  ${\overset{\mathcal{D}}\longrightarrow}$ denotes convergence in distribution, and $\tau_{1}^2=m_2-m_1^2+2\sum_{k=1}^{\infty}\gamma(k)$.
\end{theorem}
 \vskip 12pt
The proof of Theorem \ref{Th1} is given in Appendix A.3.
In Theorem 1, we do not require
$\Sigma$ to follow a diagonal matrix structure.
To derive the limiting distribution of the DLRT statistic under a general covariance matrix structure, we impose the mixing condition (C1) which implies a weak dependence structure in the data.
Specifically, noting that $T_1=\sum_{j=1}^{p} U_{nj}$,  if the autocorrelation function of
$\{U_{n1}, \ldots, U_{np}\}$ decays rapidly as the lag increases,
$T_1$ will converge to the standard normal distribution under appropriate centering and scaling.
Finally, we note that similar mixing conditions were also adopted in  Gregory et al. (2015) and Zhao and Xu (2016).
{ The asymptotic variance of $p^{1/2}T_1$, $\tau_1^2$, depends on the autocovariance sequence $\{U_{n1}, U_{n2}, \ldots\}$ and is unknown. To establish the null distribution in practice, we need an estimate, $\widehat{\tau}_{1}^2$, to replace $\tau_{1}^2$.
In spectrum analysis, under the condition (C2), we note that}
$\sum_{k=-\infty}^{\infty}\gamma(k)=2 \pi f(0)$, where $f(w)$ is a spectral density function  defined as $f(w)=(2\pi)^{-1}\sum_{k=-\infty}^{\infty}e^{iwk}\gamma(k)$ for $w \in [-\pi,\pi]$. Therefore, we only need an estimate of $f(0)$.

The estimation of $f(w)$ has been extensively studied
\citep[e.g.,][]{buhlmann1996,paparoditis2012}. The traditional kernel estimator with a lag-window form is defined as
\begin{align*}
\widehat{f}(w)=(2\pi)^{-1}\sum_{k=-\infty}^{\infty}e^{iwk} \lambda(k/h)\widehat{\gamma}(k),
\end{align*}
where $\widehat{\gamma}(k)=
p^{-1}\sum_{j=1}^{p-k}(U_{nj}-\widetilde{T}_1)(U_{n(j+k)}-\widetilde{T}_1)$ is the sample autocovariance and $\widetilde{T}_1=T_1/p$.
We apply the Parzen window \citep{Parzen1961} %\citep{Parzen1961,Priestley1962}
to determine the lag-window $\lambda(x)$ throughout the paper, where $\lambda(x)=1-6x^2-6 |x|^3$ if $|x|<1/2$, and  $\lambda(x)=2(1-|x|)^3$ if $1/2\leq |x| <1$,
and $\lambda(x)=0$ if $|x|\geq 1$.
Finally, we estimate  $\tau_1^2$ as
\begin{align*}
\widehat{\tau}_{1}^2=2 \pi \widehat{f}(0)= 2\sum_{0<k \leq h}\lambda(k/h)\widehat{\gamma}(k)+\gamma(0),
\end{align*}
where $h$ is the lag-window size, and
$\gamma(0)={\rm Var}(U_{nj})= m_2-m_1^2$.

\begin{corollary}\label{Th2}
 Let $\bm{X}_{1},\ldots,\bm{X}_{n}$ be i.i.d. random vectors from $N_{p}(\bm{\mu},\Sigma)$ and assume that $\Sigma$ is a diagonal matrix.
 Under the null hypothesis, we have the following asymptotic results:
\begin{enumerate}
  \item[$({a})$] For any fixed $n\geq2$, $({T}_1-p m_1)/{\sqrt{p(m_2-m_1^2)}}~{\overset{\mathcal{D}}\longrightarrow}~ N(0,1) \mathrm{~as~} p \to \infty$.
   \item[$({b})$] If $p$ increases at such a rate that  $p=o(n^{2k})$, then for the given positive integer $k< \nu_1/2$,
\begin{align*}
({T}_1- p \xi_k)/{\sqrt{2p}}~{\overset{\mathcal{D}}\longrightarrow}~N(0,1) \mathrm{~~as~~} (n,p) \to \infty,
\end{align*}
where $\xi_k={n}\{{a_1} - {a_2}/{2} + \cdots +(-1)^{k+1}{a_k}/{k}\}$ and $a_k=
\prod_{i=1}^{k}\{(2i-1)/(\nu_1-2i)\}$.
\end{enumerate}
\end{corollary}

The proof of Corollary \ref{Th2} is given in  Appendix A.4.
This corollary defines asymptotic normality of the DLRT statistic
for two scenarios under the diagonal covariance matrix assumption: the result from (a) establishes the asymptotic null distribution when $n$ is fixed but $p$ is large, and the result from (b) establishes the asymptotic null distribution when $n$ and $p$ are both large.

\subsection{Statistical power}\label{power.1}
To derive the asymptotic power of the proposed DLRT statistic
for the one-sample test,  we consider the local alternative
\begin{align}\label{alternative1}
\bm{\mu} -\bm{\mu}_0 = {\bm{\delta}_1}/{\sqrt{n}},
\end{align}
where $\bm{\delta}_1=(\delta_{11},\ldots,\delta_{1p})^T$. Assume that $\bm{\Delta}_1=(\Delta_{11},\ldots,\Delta_{1p})^T=
(\delta_{11}/\sigma_{11},\ldots,\delta_{1p}/\sigma_{pp})^T$, with all of the components uniformly bounded such that
\begin{align}\label{alternative1.1}
\Delta_{1j}^2 \leqslant M_0, ~{\rm{for}}~ j=1,\ldots,p,
\end{align}
where $\sigma_{jj}^2$ are the diagonal elements of $\Sigma$, and  $M_0$  is a constant  independent of $n$ and $p$.
Then we have the following theorem.

\vskip 12pt
\begin{theorem}\label{power.1}
Let $\bm{X}_{1},\ldots,\bm{X}_{n}$ be i.i.d. random vectors from $N_{p}(\bm{\mu},\Sigma)$ and assume that $p$ increases at such a rate that $p=o(n^{2})$. Let  $z_{\alpha}$ be the upper $\alpha$th percentile such that $\Phi(z_{\alpha})=1-\alpha$, where $\Phi(\cdot)$ is the cumulative distribution function of the standard normal distribution. If the sequence
$\{ U_{nj} \}$ is stationary and satisfies conditions (C1) and (C2), then under the local alternative (\ref{alternative1}) and condition (\ref{alternative1.1}), the asymptotic power of the level $\alpha$ test  is
$$\beta({T}_1)=1-\Phi\bigg(z_{\alpha}-\frac{\bm{\Delta}_1^T \bm{\Delta}_1/\sqrt{p}}{\sqrt{\tau_{1}^2}}\bigg)
\mathrm{~~as~~} (n,p) \to \infty,$$
and hence $\beta({T}_1)\to 1$ if $\sqrt{p}=o\big(\sum_{j=1}^{p}\delta_{1j}^2/\sigma_{jj}^2\big)$,
and $\beta({T}_1)\to \alpha$ if $\sum_{j=1}^{p}\delta_{1j}^2/\sigma_{jj}^2=o(\sqrt{p})$.
\end{theorem}
 \vskip 12pt
The proof of Theorem \ref{power.1} is given in Appendix A.5.
If the true mean differences are dense but small  such as the standardized signals  $(\mu_{1j}-\mu_{0j})/\sigma_{jj}=\delta_0 p^{-1/2}$
with the constant $\delta_0>0$, then the asymptotic  power will increase towards 1 as $(n,p) \to \infty$.

\section{Two-Sample Test}
In this section, we consider the two-sample test for  mean vectors with equal  covariance matrices.
Let $\bm{X}_i=(X_{i1},X_{i2},\ldots,X_{ip})^T$, $i=1,\ldots,n_1$, be i.i.d. random vectors from $N_{p}(\bm{\mu}_1,\Sigma)$,
and $\bm{Y}_k=(Y_{k1},Y_{k2},\ldots,Y_{kp})^T$, $k=1,\ldots,n_2$, be i.i.d. random vectors from $N_{p}(\bm{\mu}_2,\Sigma)$,
where $\bm{\mu}_1$ and $\bm{\mu}_2$ are two population mean vectors and $\Sigma$ is the common covariance matrix.
For ease of notation, let  $N=n_1+n_2$ and  assume that  $\lim_{N \to  \infty}n_1 /N=c \in (0,1)$. Let also
$\bm{\bar{X}}=\sum_{i=1}^{n_1}\bm{X}_i/n_1$ and $\bm{\bar{Y}}=\sum_{k=1}^{n_2}\bm{Y}_k/n_2$ be two sample mean vectors, and
\begin{equation*}
S=\frac{1}{N-2}
\Big\{\sum_{i=1}^{n_1}(\bm{X}_i-\bm{\bar{X}})(\bm{X}_i-\bm{\bar{X}})^T+
\sum_{k=1}^{n_2}(\bm{Y}_k-\bm{\bar{Y}})(\bm{Y}_k-\bm{\bar{Y}})^T \Big\}.
\end{equation*}
 be  the pooled sample covariance matrix.

In the two-sample case, we test the hypothesis
\begin{equation}\label{test1}
{H}_0: \bm{\mu}_1=\bm{\mu}_2 \text{~~versus~~}
 H_1: \bm{\mu}_1 \neq \bm{\mu}_2.
\end{equation}
In Appendix B.1,  we show that the DLRT statistic for hypothesis (\ref{test1}) is
\begin{align} \label{stat.2}
T_2
&= N\sum_{j=1}^{p}\mathrm{log}
\Bigl\{1+\frac{n_1n_2}{N(N-2)}
\frac{(\bar{X}_j-\bar{Y}_{j})^2}{s_{j,\mathrm{pool}}^2} \Bigr\}
= N\sum_{j=1}^{p}\mathrm{log}
\Bigl(1+\frac{t_{Nj}^2}{\nu_2}\Bigl),
\end{align}
where $t_{Nj}=\sqrt{n_1n_2/N}(\bar{X}_j-\bar{Y}_{j})/s_{j,\mathrm{pool}}$
are the standard $t$-statistics for the two-sample case with
 $\nu_2=N-2$  degrees of freedom, and
 $s_{j,\mathrm{pool}}^2$ are the pooled sample variances, i.e., the diagonal elements of $S$.

For ease of notation,  let  $V_{Nj}= N\log(1+t_{Nj}^2/\nu_2)$ for $j=1,\ldots, p$. The following theorem establishes the asymptotic null distribution of the DLRT statistic for the two-sample case  under centering and scaling.
\vskip 12pt
\begin{theorem}\label{Th4}
Let $\{\bm{X}_{i}\}_{i=1}^{n_1}$ and $\{\bm{Y}_{k}\}_{k=1}^{n_2}$
be i.i.d. random vectors from $N_{p}(\bm{\mu}_1,\Sigma)$ and
$N_{p}(\bm{\mu}_2,\Sigma)$, respectively.
If the sequence $\{ V_{Nj} \}$  is stationary and satisfies conditions (C1) and (C2), then under the null hypothesis, we have for any fixed $N \geq 4$,
\begin{align*}
\frac{{T}_2-p G_1}{\tau_{2}\sqrt{p}}
~{\overset{\mathcal{D}}\longrightarrow}~ N(0,1)
{\rm~~as~~} p \rightarrow \infty
\end{align*}
 where $\tau_{2}^2=G_2-G_1^2+2\sum_{k=1}^{\infty}\gamma(k)$, with
$G_1= ND(\nu_2)$
 and $G_2 ={N^2\{D^2(\nu_2)-2D'(\nu_2)\}}$.
\end{theorem}

 \vskip 12pt
The proof of Theorem \ref{Th4} is given in Appendix B.2.
{By imposing conditions (C1) and (C2) on the sequence $\{ V_{Nj} \}$, Theorem \ref{Th4} also does not require the assumption that each of the covariance matrices follows a diagonal matrix structure.}
Similar to  the one-sample case, a consistent estimator for $\tau_{2}^2$ is given as
\begin{align*}
\widehat{\tau}_{2}^2= 2 \sum_{0<k\leq h}\lambda(k/h)\widehat{\gamma}(k)+\gamma(0),
\end{align*}
where $\lambda(x)$ is the Parzen window, $h$ is the lag-window size,
$\gamma(0)={\rm Var}(V_{Nj})= G_2-G_1^2$, and
$\widehat{\gamma}(k)=
p^{-1}\sum_{j=1}^{p-k}(V_{Nj}-\widetilde{T}_2)(V_{N(j+k)}-\widetilde{T}_2)$
 is the sample autocovariance for $\{V_{Nj}, j=1,2,\ldots,p \}$ and
 $\widetilde{T}_2=T_2/p$.

\begin{corollary}\label{equal.th5}
Let $\{\bm{X}_{i}\}_{i=1}^{n_1}$ and $\{\bm{Y}_{k}\}_{k=1}^{n_2}$
be i.i.d. random vectors from $N_{p}(\bm{\mu}_1,\Sigma)$ and
$N_{p}(\bm{\mu}_2,\Sigma)$, respectively, and assume that $\Sigma$ is a diagonal matrix.
Under the null hypothesis, we have the following asymptotic results:
\begin{enumerate}
  \item[$(a)$] For any fixed $N \geq 4$,
 $({{T}_1-pG_1})/\sqrt{p(G_2-G_1^2)}{~\overset{\mathcal{D}}\longrightarrow~} N(0,1)\mathrm{~as~} p \to \infty.$
  \item[$(b)$] If $p$ increases at such a rate that  $p=o(N^{2k})$,
      then for the given positive integer $k< \nu_2/2$,
\begin{align*}
({{T}_2-p \eta_k})/\sqrt{2p}
~{\overset{\mathcal{D}}\longrightarrow}~ N(0,1) \mathrm{~~as~~} (N,p) \to \infty,
\end{align*}
where $\eta_k={N}\{{b_1} - {b_2}/{2} + \cdots +(-1)^{k+1}{b_k}/{k}\}$ and $b_k=
\prod_{i=1}^{k}\{(2i-1)/(\nu_2-2i)\}$.
\end{enumerate}
\end{corollary}

The proof of Corollary \ref{equal.th5} is given in Appendix  B.3.
This corollary defines asymptotic normality of the DLRT statistic
for two scenarios under the diagonal covariance matrix assumption: the result from (a)  establishes the asymptotic null distribution when $N$ is fixed but $p$ is large, and the result from (b) establishes the asymptotic null distribution when $N$ and $p$ are both large.

When $\bm{\mu}_1 \neq \bm{\mu}_2$, we consider  the local alternative
\begin{align}\label{alternative2}
\bm{\mu}_1 -\bm{\mu}_2 = \sqrt{\frac{N}{n_1n_2}}\bm{\delta}_2,
\end{align}
where $\bm{\delta}_2=(\delta_{21},\ldots,\delta_{2p})^T$.
We assume that $\bm{\Delta}_2=(\Delta_{21},\ldots,\Delta_{2p})^T=
(\delta_{21}/\sigma_{11},\ldots,\delta_{2p}/\sigma_{pp})^T$, with all of the components uniformly bounded such that
\begin{align}\label{alternative2.1}
\Delta_{2j}^2 \leqslant M_1, ~{\rm{for}}~ j=1,\ldots,p,
\end{align}
where $\sigma_{jj}^2$ are the diagonal elements of $\Sigma$, and  $M_1$  is a constant independent of $N$ and $p$.
The following theorem establishes the asymptotic power of our proposed  DLRT statistic  for the two-sample test.

\begin{theorem}\label{power.2}
Let $\{\bm{X}_{i}\}_{i=1}^{n_1}$ and $\{\bm{Y}_{k}\}_{k=1}^{n_2}$
be i.i.d. random vectors from $N_{p}(\bm{\mu}_1,\Sigma)$ and
$N_{p}(\bm{\mu}_2,\Sigma)$, respectively.
Assume that $p$ increases at such a rate that $p=o(N^{2})$.
If the sequence $\{V_{Nj},j=1,2,\ldots\}$ is stationary and satisfies conditions (C1) and (C2), then  under the local alternative (\ref{alternative2}) and condition (\ref{alternative2.1}), the asymptotic power of the level $\alpha$ test is
\begin{align*}
\beta(T_2)=1-\Phi \bigg(z_{\alpha}-\frac{\bm{\Delta}_2^T \bm{\Delta}_2/\sqrt{p}}{\sqrt{\tau_{2}^2}}\bigg)
\mathrm{~~as~~} (N,p) \to \infty,
\end{align*}
and hence, $\beta(T_2) \to 1$ if $\sqrt{p}=o\big(\sum_{j=1}^{p}\delta_{2j}^2/\sigma_{jj}^2\big)$, and
        $\beta(T_2) \to \alpha$ if $\sum_{j=1}^{p}\delta_{2j}^2/\sigma_{jj}^2=o(\sqrt{p})$.
\end{theorem}

\section{Monte Carlo Simulation Studies}
In this section, we carry out simulations to evaluate the performance of our DLRT method. For ease of presentation, we consider the proposed DLRT test for the two-sample case only. We compare DLRT with five existing tests from the aforementioned three categories:
one  unscaled Hotelling's test including  the CQ test from \citet{chen2010},
one  regularized Hotelling's test including  the RHT test from \citet{chen2011}, and
two diagonal Hotelling's tests including the SD test from \citet{srivastava2008},
and the GCT test from \citet{Carroll2015}.
\citet{Carroll2015} considered two different versions of the GCT test with centering corrections that allowed the dimension to grow at either a moderate or large order of the sample size, which are denoted as ${\rm GCT}_{\rm md}$ and ${\rm GCT}_{\rm lg}$, respectively. The lag-window size throughout the simulations is $h = 5$.

\subsection{Normal data}\label{normal.data}
In the first simulation, we generate $\bm{X}_1,\ldots,\bm{X}_{n_1}$
from $N_{p}(\bm{\mu}_1,\Sigma)$, and $\bm{Y}_1,\ldots,\bm{Y}_{n_2}$
from $N_{p}(\bm{\mu}_2,\Sigma)$. For simplicity, let $\bm{\mu}_1=\bm{0}$. Under the alternative hypothesis, we assume that the first $p_0$ elements in $\bm{\mu}_2$ are nonzero, where $p_0=\beta p$ with  $\beta \in [0,1]$ being the tuning parameter that controls  the signal sparsity. When $\beta=0$, the null hypothesis holds. The common covariance matrix is $\Sigma=D^T R D$, where
$R$ is the correlation matrix and $D$ is a diagonal matrix such that $D=\mathrm{diag}(\sigma_{11},\sigma_{22}, \ldots,\sigma_{pp})$.
To account for  the heterogeneity of variances, $\sigma_{11}^2,\ldots,\sigma_{pp}^2$  are randomly sampled from the
scaled chi-square distribution $\chi_{5}^2/5$.
For the dependence structure in the matrix $R$, we consider the following three scenarios:
\begin{enumerate}
  \item[$({\rm a})$] Independent (IND) structure:  $R$ is the $p \times p$ identity matrix.
  \item[$({\rm b})$] Short range dependence (SRD) structure: $R=(\rho^{|i-j|})_{p \times p}$ follows the first-order autoregressive structure, in which the correlation among the observations decay exponentially with distance. We consider $\rho=0.3$  or $0.6$ to represent two different levels of correlation.
\item[$({\rm c})$]  Long range dependence (LRD) structure: We follow the same setting as in \citet{Carroll2015}. Specifically, we consider the $(i,j)$th element of $R$ as $r_{ij}=[(k+1)^{2 H}+(k-1)^{2 H}-2k^{2 H}]/2$ with $k=|j-i|$, and the self-similarity parameter as $H=0.625$.
\end{enumerate}

For the power comparison, we set the $j$th nonzero component in $\bm{\mu}_2$ as $\mu_{2j}=\theta \sigma_{jj}, j=1,\ldots, p_0$, where $\theta$ is the effect size of the corresponding component. The other parameters are set as
$(n_1,n_2,\theta)  \times  p =\{(3,3,0.5) {~\rm or~} (5,5,0.5) {~\rm or~} (15,15,0.25)\}
\times  \{100 {~\rm or~} 500\}$, respectively.

\begin{figure}[!tbp]
    \centering
    {\includegraphics[width=15cm,height=14cm]{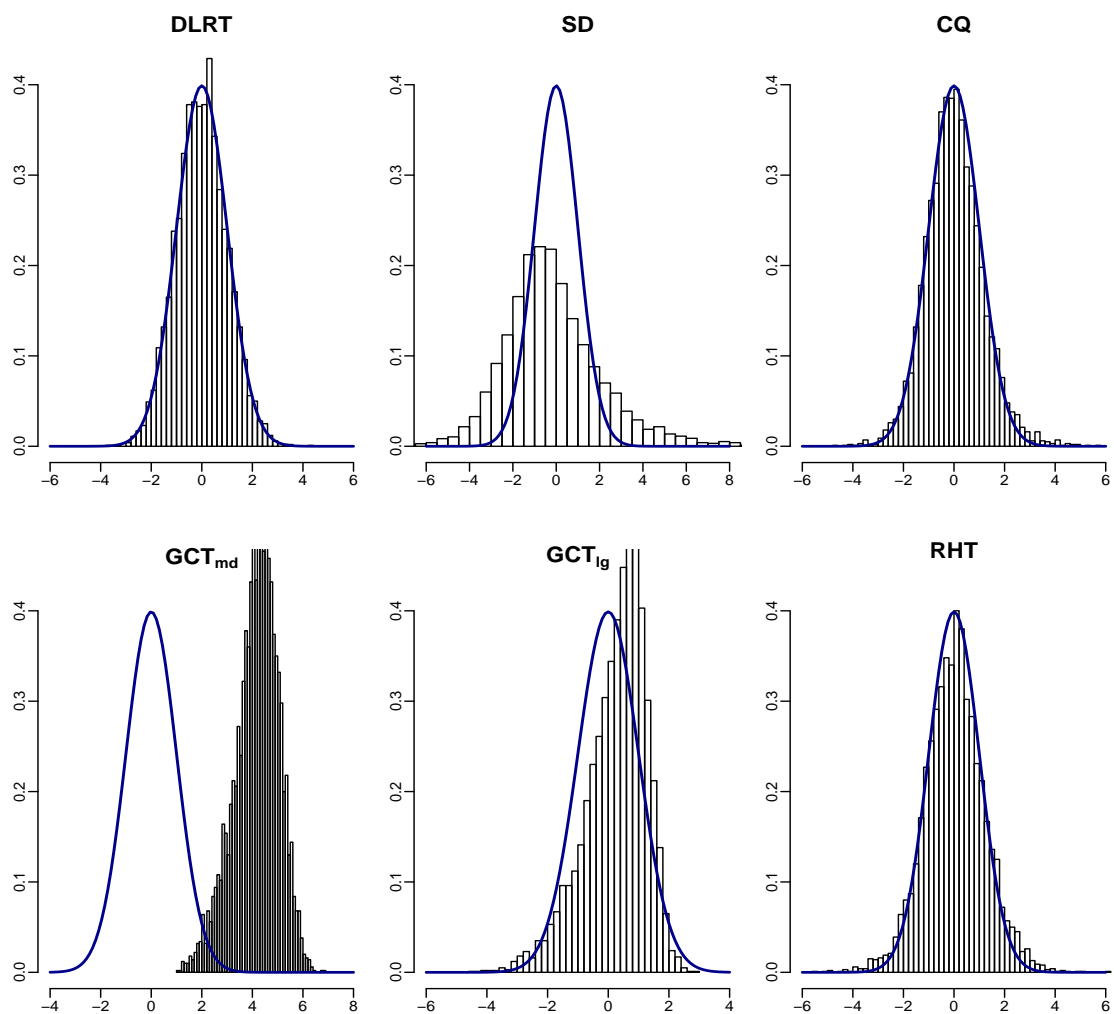}}
\caption{\label{CLT.appro} Comparison with the standard normal distribution under the null hypothesis for the DLRT, SD, CQ, ${\rm GCT}_{\rm md}$,${\rm GCT}_{\rm lg}$ and RHT tests with $n_1=n_2=3$ and $p=500$. The histograms are based on 5000 simulations.
 }
\end{figure}

Figure \ref{CLT.appro} shows the simulated null distributions of the DLRT, SD, ${\rm GCT}_{\rm md}$, ${\rm GCT}_{\rm lg}$, CQ and RHT tests under the independent structure, when the sample size is small (e.g., $n_1=n_2=3$)
and the dimension is large. The histograms are based on 5000 simulations. For comparison, their limiting distributions are also plotted.
However, the null distributions of the other three tests, and especially the ${\rm GCT}_{\rm md}$  test, are either skewed or shifted away from the standard normal distribution.

We summarize the type I error rates from the simulations for each of the six tests, with different sample sizes and dependence structures, in Table \ref{TypeI.1}. When the variables are uncorrelated or weakly correlated with each other, the type I error rates of DLRT are closer to the nominal level ($\alpha=0.05$) than the other five tests under most settings.
In addition, DLRT provides a more stable test statistic and better control over the type I error rate when the sample size is not large; the SD, ${\rm GCT}_{\rm lg}$, CQ and RHT tests have inflated type I error rates when the sample size is relatively small (e.g., $n_1=n_2=3$).
%the SD and ${\rm GCT}_{\rm md}$ tests exhibit substantially inflated type I errors when the sample size is %relatively small %(e.g., $n_1=n_2=3$).
The  ${\rm GCT}_{\rm md}$ test in particular fails to keep the type I error rate within the nominal level under each setting, and performs more poorly when the sample size is small and the dimension is large.
Therefore, we exclude the ${\rm GCT}_{\rm md}$ test from the following power comparison.

\begin{figure}[!htbp]
    \centering
    \subfigure[$n_1=n_2=5, p=100$]
    {\includegraphics[width=16cm,height=9.5cm]{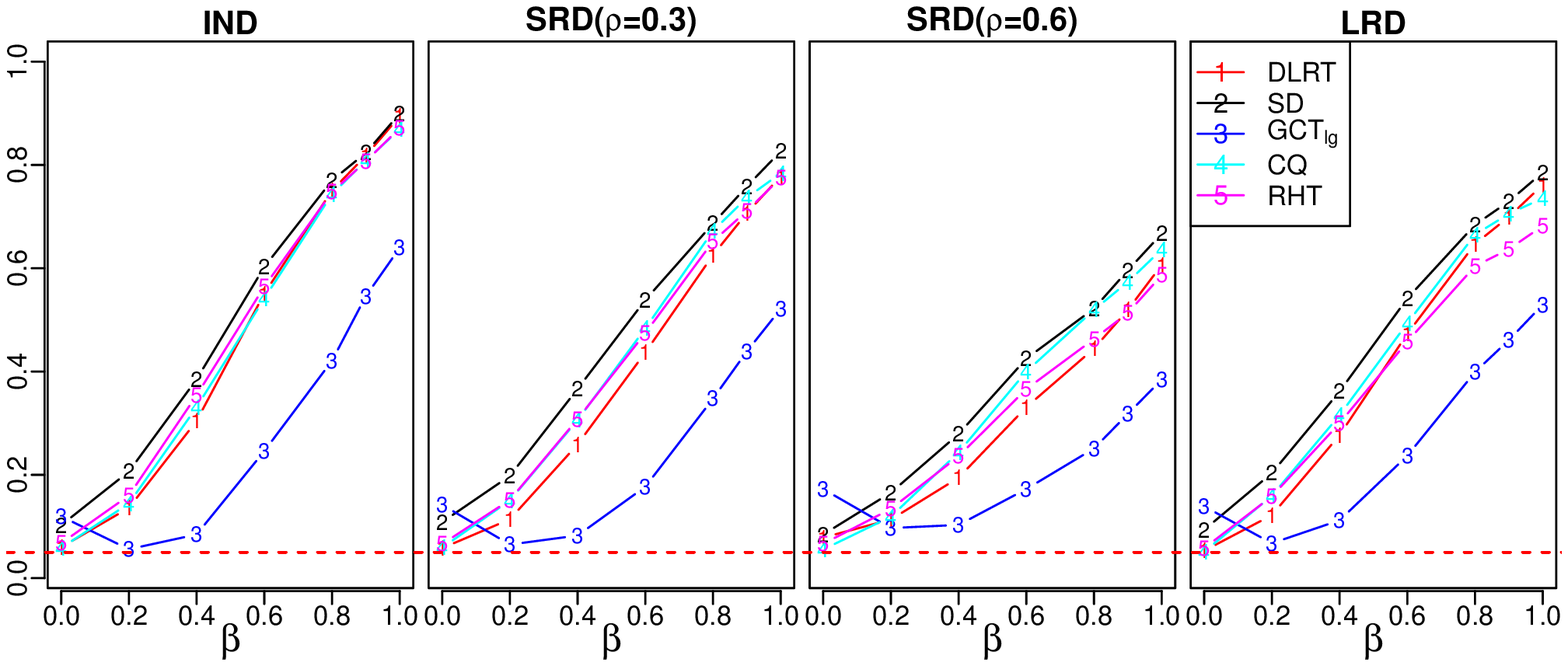}}\\
    \subfigure[$n_1=n_2=15, p=500$]
    {\includegraphics[width=16cm,height=9.5cm]{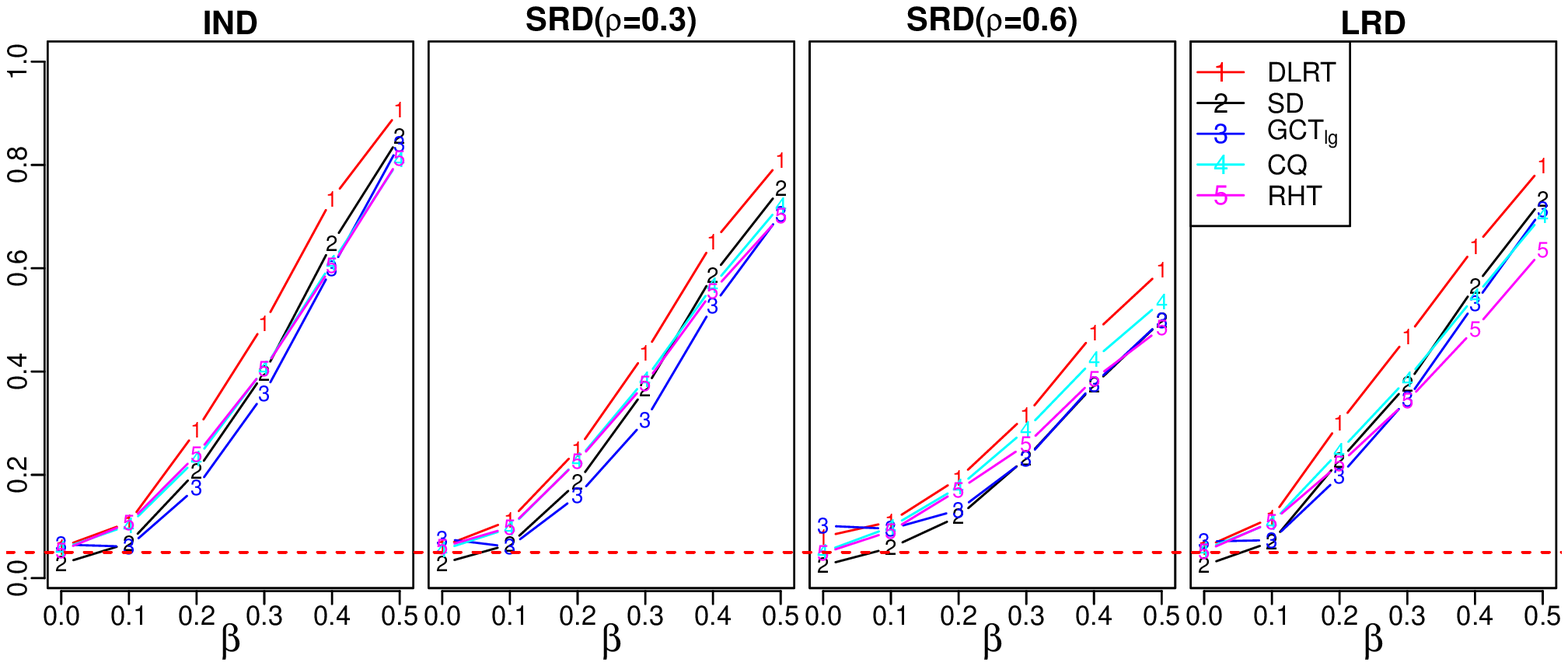}}
    \caption{\label{power.n=5} Power comparisons among the DLRT, SD,  ${\rm GCT}_{\rm lg}$, CQ and RHT tests with ($n_1=n_2=5, p=100$) or ($n_1=n_2=15, p=500$), respectively.
    The horizontal dashed red lines represent the significance level of $\alpha=0.05$.  The results are based on 2000 simulations with data from the normal distribution.}
\end{figure}

Figure \ref{power.n=5} presents the simulated  power of the DLRT, SD, ${\rm GCT}_{\rm lg}$,
CQ and RHT tests at the significance level $\alpha=0.05$. When the dimension is low (e.g., $p=100$),
the DLRT, CQ  and RHT tests are able to control the type I error rates well, whereas the SD and  ${\rm GCT}_{\rm lg}$ tests suffer from inflated type I error rates.
In particular for the ${\rm GCT}_{\rm lg}$ test, it exhibits a relatively low power when the sample size is small.
This coincides with the findings in Figure \ref{CLT.appro}.
As the dimension is large and the sample size is not small, the DLRT, SD, CQ and RHT  tests control the type I error rate  close to the nominal level, whereas, the ${\rm GCT}_{\rm lg}$ test still fails.
DLRT also provides a higher power in most settings.
To conclude, DLRT  performs comparably to the existing tests for normal data.

\subsection{Heavy-tailed data}
To evaluate the robustness of  DLRT, we also conduct simulations with heavy-tailed data. Following \citet{Carroll2015}, the data are generated based on a ``double" Pareto distribution with parameters $a$ and $b$.
The algorithm is as follows:
\begin{enumerate}
\item[$({\rm i})$] Generate two independent random variables $U$ and $V$, where $U$ is from the Pareto distribution with  the cumulative distribution function $F(x)=1-(1+x/b)^{-a}$ for $x\geq0$, and  $V$ is a binary random variable with $P(V=1)=P(V=-1)=0.5$. Then $Z=UV$ follows the double Pareto distribution with parameters $a$ and $b$.
\item[$({\rm ii})$] Generate random vectors $\{\bm{X}_i^{(0)}=(x_{i1},\ldots,x_{ip})^T\}_{i=1}^{n_1}$, and random vectors $\{ \bm{Y}_k^{(0)}=(y_{k1},\ldots,y_{kp})^T \}_{k=1}^{n_2}$, where all the components of $\bm{X}_i^{(0)}$ and $\bm{Y}_k^{(0)}$ are sampled independently from the double Pareto distribution with parameters  $a=16.5$ and $b=8$.
  \item[$({\rm iii})$] Let ${\textstyle {\bm{X}_i=\bm{\mu}_1+\Sigma^{1/2}\bm{X}_i^{(0)}/c_0} }$ and ${\textstyle\bm{Y}_k=\bm{\mu}_2+\Sigma^{1/2}\bm{Y}_k^{(0)}/c_0}$,
      where $c_0^2={512}/{899}$ is the variance of the double Pareto  distribution with $a=16.5$ and $b=8$, and $\Sigma=D^T R D$  with $D=\mathrm{diag}(\sigma_{11}, \ldots, \sigma_{pp})$.   Consequently, $\bm{X}_i$  and $\bm{Y}_k$ have a common correlation matrix $R$.
\end{enumerate}
For the matrix $R$, we also consider three scenarios: (a) the IND structure, (b) the SRD structure, and (c) the LRD structure. In each scenario, the generating algorithms for $\bm{\mu}_1$, $\bm{\mu}_2$ and $\Sigma$ follow the simulation procedure described in Section 4.1.
{The parameters used in the algorithms} are  $(n_1,n_2,\theta)  \times  p =\{(5,5,0.5) {\rm ~or~} (15,15,0.25)\}\times  \{100
{\rm ~or~} 500\}$, respectively.

\begin{figure}[!htbp]
    \centering
    \subfigure[$n_1=n_2=5, p=100$]
    {\includegraphics[width=16cm,height=9.5cm]{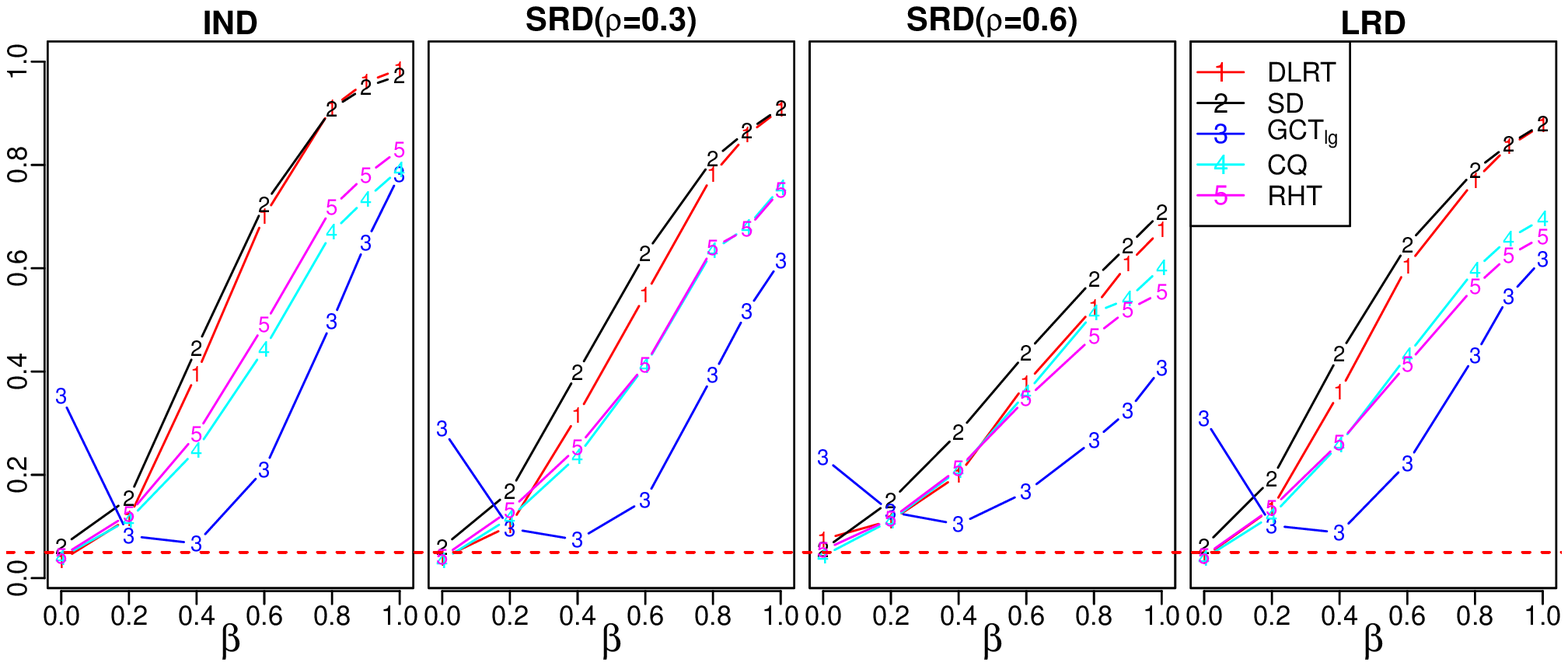}}\\
    \subfigure[$n_1=n_2=15, p=500$]
    {\includegraphics[width=16cm,height=9.5cm]{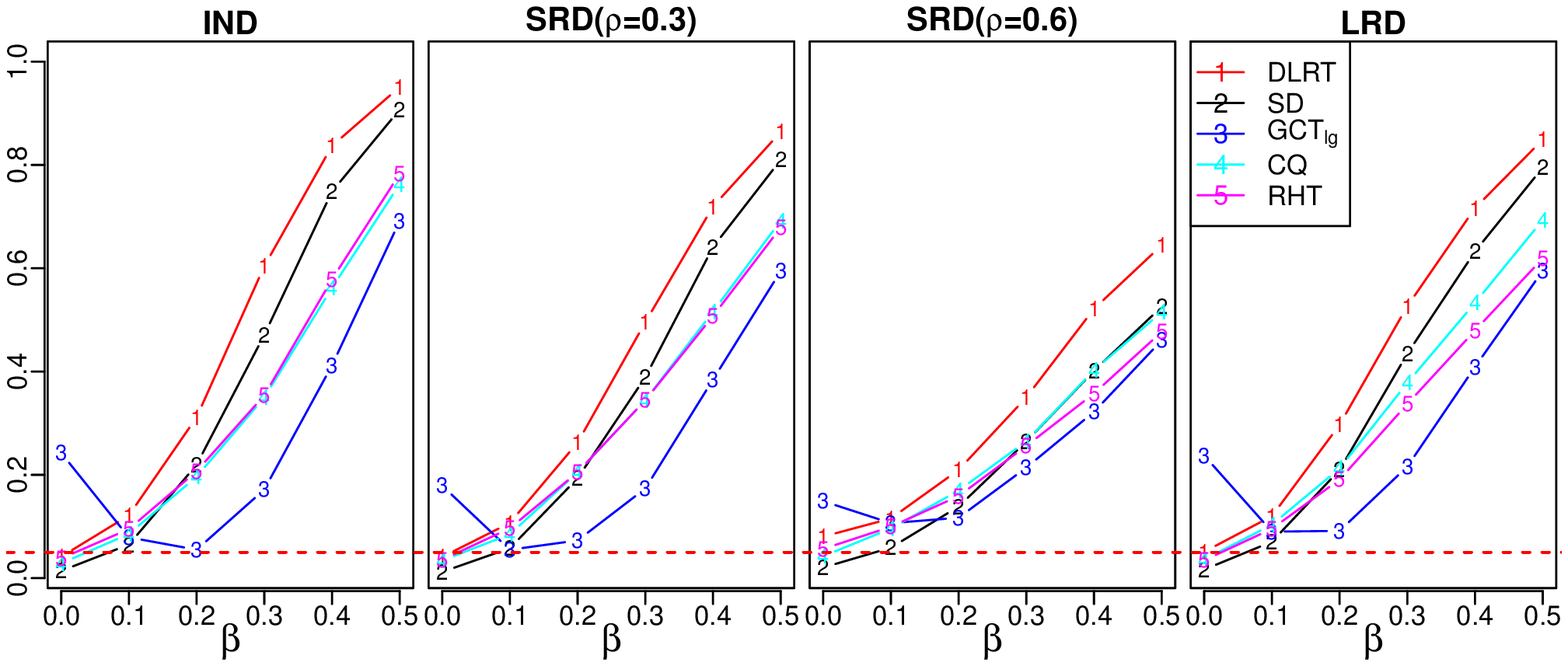}}
     \caption{\label{ultra.power.n=5} Power comparisons among the DLRT, SD, ${\rm GCT}_{\rm lg}$, CQ and RHT tests with ($n_1=n_2=5$, $p=100$) or ($n_1=n_2=15$, $p=500$), respectively.
     The  horizontal dashed red lines represent  the significance level of $\alpha=0.05$. The results are  based on 2000 simulations with data from a heavy-tailed distribution.}
\end{figure}

Figure \ref{ultra.power.n=5} presents the simulation results for  the five tests with heavy-tailed data at the significance level $\alpha=0.05$.
When the dimension is large and the sample size is small,
the DLRT, SD and RHT tests control the type I error rate well,
whereas the ${\rm GCT}_{\rm lg}$ test exhibits a substantially inflated type I error rate and a low power for detection. One possible explanation is that the ${\rm GCT}_{\rm lg}$ statistic involves the estimation of high order moments which leads to instability when the sample size is small.
DLRT is again more powerful than the CQ and RHT tests in most settings.
In summary, it is evident that the DLRT test provides a more robust performance with heavy-tailed data than the existing five tests, especially when the dimension is large.

\section{Brain Cancer Data Analysis}
In this section, we apply DLRT to a data set from The Cancer Genome Atlas (TCGA). This data set contains the copy number measurements from
genomic locations of the probes on chromosomes in 92 long-term survivors and 138 short-term survivors with a brain cancer called glioblastoma multiforme. The long-term brain cancer survivors lived for more than two years after their first diagnosis, and the short-term survivors lived for less than two years after their first diagnosis.
According to  \citet{olshen2004} and \citet{Baladandayuthapani2010},  the copy number variations between the patient groups will occur across multiple probes rather than at a single probe. That is, the signal structure is dense-but-small  rather than sparse-but-strong.
To identify the particular regions in the genome where the genes were differentially expressed,  we apply the  following tests: the DLRT, SD, ${\rm GCT}_{\rm lg}$, CQ and RHT tests. \citet{Carroll2015} separated  the whole chromosome into 26 segments of varying lengths. We focus our analysis on one segment of the $q$ arm of chromosome 1, which contains measurements of probes at 400 locations. The copy number data at 400 locations are summarized in ``chr1qseg.rda"  which is available from the R package ``highD2pop".

To compare the performance of the  tests,
we first perform the two-sample $t$-tests to screen top $p$ significant genes, and then calculate the empirical power with $p=100$, $200$ or $400$, respectively.
To determine the empirical critical values corresponding to a given nominal level $\alpha$, we bootstrap two distinct classes from the short-term survival group to compute the test statistics. Since both classes are partitioned  from the short-term survival group, the null hypothesis can
be regarded as the truth. Therefore, we repeat the procedure 10,000 times for each test method, and select the $(10,000\alpha)$th largest value of the test statistics as the empirical critical values.  To determine the empirical power, we bootstrap one class from the short-term survival  group and another class from the long-term survival group. For both classes, we consider $n_1=n_2=8$  for computing  the empirical critical values and power.

Table \ref{empirical-power} shows the empirical power of the DLRT, SD, ${\rm GCT}_{\rm lg}$, CQ and RHT tests.
We note that the DLRT test performs nearly as well as the RHT test, and it has a higher empirical power than the other three tests under all the settings.

\section{Conclusion}
In the classical low-dimensional setting,
Hotelling's $T^2$  test is an important and useful tool for testing the equality of one or two mean vectors from multivariate normal distributions.
However, this classic test may not be applicable when the dimension is larger than the sample size, as the sample covariance matrix is no longer invertible. This motivates the development of new methods to address the testing problems for high-dimensional data with a small sample size. According to how the covariance matrices are estimated, most available methods can be classified into three categories: the unscaled Hotelling's tests, the regularized Hotelling's tests, and the diagonal Hotelling's tests.

In this paper, we proposed a new test framework based on the likelihood ratio test for both one- and two-sample cases. The proposed test statistics are derived under the assumption that the covariance matrices follow a diagonal matrix structure. Our tests use the log-transformed squared $t$-statistics and provide more stable test statistics than the standard $t$-statistics when the sample size is small. Through simulation studies, we showed that DLRT is also more robust than the existing test methods when the data are heavy-tailed or weakly correlated. In other words, when the dimension is large and the sample size is small, DLRT is able to keep the type I error rate within the nominal level and, at the same time, maintains a high power for detection.

The proposed  new test assumes a natural ordering of the components in the $p$-dimensional random vector, e.g., the correlation among the components are related to their positions, and hence we can take into account the additional structure information to avoid an estimation of the full covariance matrix.
When the ordering of the components is not available, we propose to reorder the components from the sample data using some well known ordering methods before applying our proposed tests. For instance, with the best permutation algorithm in \citet{rajaratnam2013}, the strongly correlated elements can be reordered close to each other. For other ordering methods of  random variables, one may refer to, for example, \citet{gilbert1992}, \citet{wagaman2009}, and the references therein.

{When the sample size is relatively small and the correlation is very  high, our proposed tests will have slightly inflated type I error rates, especially when the dimension is also large. This is mainly because the test statistics are derived under the assumption that the covariance matrices follow a diagonal matrix  structure. When the diagonal matrix assumption is violated, the asymptotic null distributions may not follow the standard normal distribution, or the asymptotic properties may require more restrictive assumptions including a larger sample size. To overcome these limitations, future research is warranted to  improve our current version of DLRT or to derive  more accurate asymptotic distributions when the underlying assumptions are violated.}

We also note that our current paper has focused on testing high-dimensional mean vectors under the parametric setting. More recently, some nonparametric tests have also been developed
 in the literature  for the same testing problems; see, for example,  \citet{wang2015}, \citet{ghosh2016} and \citet{chakraborty2017}.

\section{Supplementary Materials}
Web Appendix referenced in Sections 2 and 3 is available
with this paper at the {\it Biometrics} website on Wiley Online Library.

\section{Acknowledgements}
The authors thank the editor, the associate editor and three reviewers for their constructive comments that have led to a substantial improvement of the paper.
Tiejun Tong's research was supported by the Hong Kong RGC Grant (No. HKBU12303918), the National Natural Science Foundation of China (No. 11671338),  the Health and Medical Research Fund (No. 04150476), and two Hong Kong Baptist University grants (FRG1/17-18/045 and FRG2/17-18/020). Marc G. Genton's research was supported by the King Abdullah University of Science and Technology (KAUST).

\linespread{1}

\begin{center}
{
\begin{table}[htbp!]
\caption{\label{TypeI.1} Type I error rates over 2000 simulations for the  DLRT, ${\rm GCT}_{\rm md}$,
${\rm GCT}_{\rm lg}$, SD, CQ, RHT tests  under three dependence structures.
The significance level is $\alpha=0.05$.  Two different correlation, $\rho=0.3$ or $0.6$, are considered for the SRD structure.}
{ {
\begin{center}
\begin{tabular}{c|c||c|c|c||c|c|c}
\hline\hline
& & \multicolumn{3}{c||}{$p=100$} & \multicolumn{3}{c}{$p=500$} \\\hline
\multirow{2}{*}{Cov}& \multirow{2}{*}{{Method}~\textbackslash{} {$(n_1,n_2)$}} & \multirow{2}{*}{$~(3,3)~$}
& \multirow{2}{*}{$~(5,5)~$}& \multirow{2}{*}{$(15,15)$}&
 \multirow{2}{*}{$~(3,3)~$}& \multirow{2}{*}{$~(5,5)~$}&
  \multirow{2}{*}{$(15,15)$}\\
&  & & & & & & \\\hline
\multirow{6}{*}{IND}
&DLRT&0.060&0.056&0.058&0.055&0.043&0.048    \\\cline{2-8}
&SD&0.403&0.111&0.044&0.342&0.038& 0.024  \\\cline{2-8}
%&PA& --- &0.087&0.051&---&0.099& 0.057   \\\cline{2-8}
&${\rm GCT}_{\rm md}$&0.608&0.233&0.063&0.983&0.922&0.151  \\\cline{2-8}
&${\rm GCT}_{\rm lg}$ &0.101&0.124&0.092&0.047&0.110&0.060\\\cline{2-8}
&CQ&0.092&0.066&0.056&0.084&0.058& 0.051  \\\cline{2-8}
&RHT&0.097&0.068&0.058&0.105&0.070& 0.053   \\ \hline
\hline
\multirow{5}{*}{SRD}
&DLRT&0.067&0.058&0.054&0.054&0.053& 0.061  \\\cline{2-8}
&SD&0.387&0.105& 0.038&0.309&0.046& 0.023  \\\cline{2-8}
%&PA&---&0.098 & 0.056&--- &0.091 & 0.054  \\\cline{2-8}
&${\rm GCT}_{\rm md}$&0.548&0.203&0.071&0.986&0.893&0.143 \\\cline{2-8}
($\rho=0.3$)
&${\rm GCT}_{\rm lg}$&0.116&0.119&0.094 &0.05&0.106&0.070 \\\cline{2-8}
&CQ&0.076&0.054&0.052&0.078&0.059& 0.051  \\\cline{2-8}
&RHT& 0.099 & 0.067& 0.058&0.098 & 0.060&0.060    \\ \hline
\hline
\multirow{5}{*}{SRD}
&DLRT&0.072&0.076&0.078&0.080&0.072&0.078  \\\cline{2-8}
&SD&0.351&0.082&0.031&0.28&0.041&0.023 \\\cline{2-8}
%&PA&---& 0.068 &  0.049 & --- & 0.095 &  0.054  \\\cline{2-8}
&${\rm GCT}_{\rm md}$&0.421&0.152&0.091&0.980&0.749&0.116 \\\cline{2-8}
$(\rho=0.6)$
&${\rm GCT}_{\rm lg}$&0.130&0.164&0.129&0.064&0.122&0.095 \\\cline{2-8}
&CQ&0.079&0.050&0.056&0.080&0.051& 0.053  \\\cline{2-8}
&RHT& 0.110 & 0.067 & 0.060 & 0.099 &0.063 & 0.048   \\\hline \hline
 \multirow{7}{*}{LRD}
&DLRT&0.061&0.065& 0.054&0.052&0.071& 0.056\\\cline{2-8}
&SD &0.383&0.112&0.034&0.332&0.054&0.025 \\\cline{2-8}
%&PA&---&0.087& 0.051&---&0.095 & 0.054   \\\cline{2-8}
&${\rm GCT}_{\rm md}$&0.594&0.222& 0.067&0.983&0.913& 0.143 \\\cline{2-8}
&${\rm GCT}_{\rm lg}$&0.104&0.152&0.095&0.047&0.109& 0.077 \\\cline{2-8}
&CQ&0.092&0.058&0.052&0.082&0.062& 0.057  \\\cline{2-8}
&RHT& 0.108&0.058&0.060&0.102 & 0.063& 0.052   \\ \hline \hline
\end{tabular}\end{center}
} }
\end{table}
}
\end{center}

\begin{center}
{
\begin{table}[htbp!]
\caption{\label{empirical-power}
Empirical power for testing the equality of gene expressions in the TCGA data, when $p=100, 200$ or $400$.
The nominal level is $\alpha$ and the sample sizes of the two classes are $n_1=n_2=8$.}
{
\begin{center}
\begin{tabular}{c|c||c|c|c|c|c}
\hline\hline
& &DLRT&SD&${\rm GCT}_{\rm lg}$&CQ&RHT\\  \hline
\multirow{3}{*}{$\alpha=0.05$}
&$p=100$&0.165&0.121&0.082&0.143&0.189\\ \cline{2-7}
&$p=200$&0.158&0.096&0.094&0.116&0.194\\ \cline{2-7}
&$p=400$ &0.144&0.081&0.119&0.106&0.136\\ \hline\hline
\multirow{3}{*}{$\alpha=0.10$}
&$p=100$&0.274&0.201&0.165&0.233&0.345\\ \cline{2-7}
&$p=200$&0.261&0.157&0.190&0.192&0.354\\ \cline{2-7}
&$p=400$&0.258&0.154&0.212&0.181&0.295\\ \hline
\end{tabular}\end{center}
}
\end{table}
}
\end{center}

\linespread{1.25}

\clearpage

\noindent{\huge  Web-based Supplementary Materials  for ``Diagonal Likelihood Ratio Test for  Equality of Mean Vectors in High-Dimensional Data"}

\begin{appendices}
%%\appendix
\renewcommand\thesection{{A}}{}
\section{DLRT for One-Sample Test}
\renewcommand\thesection{{A}}{}
\subsection{Derivation of Formula (4)}\label{stat1}

Under the assumption  $\Sigma={\text{diag}}(\sigma_{11}^2,\sigma_{22}^2,\ldots,\sigma_{pp}^2)$,
we can write the likelihood function as
$$L(\bm{\mu}, \Sigma)=\prod_{j=1}^{p} L_j(\mu_j, \sigma_{jj}^2),$$
where $L_j(\mu_j, \sigma_{jj}^2)=(2 \pi)^{-{n \over 2}}{(\sigma_{jj}^2)}^{-{n \over 2}}
e^{-\frac{1}{2}\sum_{i=1}^{n}(X_{ij}-\mu_j)^2/ \sigma_{jj}^2 }$ is the likelihood function for the $j$th component with data $X_{1j},\ldots,X_{nj}$.
Deriving  the maximum likelihood estimator of $L(\bm{\mu}, \Sigma)$ is equivalent to finding the maximum likelihood estimators of  $L_j(\mu_j, \sigma_{jj}^2)$ for $j=1,2,\ldots,p$, respectively.

It is known that the maximum of $L_j(\mu_j, \sigma_{jj}^2)$ is achieved when
$\widehat{\mu}_j=\bar{X}_j=\sum_{i=1}^{n}X_{ij}/n$ and
$s_j^2=\sum_{i=1}^{n}(X_{ij}-\bar{X}_j)^2/(n-1)$.
Hence, under the alternative hypothesis,
\begin{align*}
\underset{\bm{\mu},\Sigma}{\mathrm{max}}L(\bm{\mu}, \Sigma)
&=(2 \pi)^{-{np\over 2}} e^{-{np\over 2}} \prod_{j=1}^{p}\Bigl\{\frac{1}{n}
\sum_{i=1}^{n}(X_{ij}-\bar{X}_j)^2 \Bigr\}^{-{n\over 2}}.
\end{align*}
Similarly, under the null hypothesis,
\begin{align*}
\underset{\Sigma}{\mathrm{max}}L(\bm{\mu}_0, \Sigma)
&=(2 \pi)^{-{np\over 2}} e^{-{np\over 2}} \prod_{j=1}^{p}\Bigl\{\frac{1}{n}
\sum_{i=1}^{n}(X_{ij}-\mu_{0j})^2 \Bigr\}^{-{n\over 2}}.
\end{align*}
From the above results, the likelihood ratio test statistic is given as
\begin{align*}
\Lambda_n&=\frac{\underset{\Sigma}{\mathrm{max}} L(\bm{\mu}_0,\Sigma)}
{\underset{\bm{\mu},\Sigma}{\mathrm{max}} L(\bm{\mu},\Sigma)}
=\frac{\prod_{j=1}^{p} \bigl\{ \sum_{i=1}^{n}(X_{ij}-\bar{X}_j)^2   \bigr\}^{{n\over 2}}   }
{\prod_{j=1}^{p} \bigl\{ \sum_{i=1}^{n}(X_{ij}-\mu_{0j})^2   \bigr\} ^{{n\over 2}} }.
\end{align*}
Further, we have
\begin{align*}
\Lambda_n ^{2/n}=\prod_{j=1}^{p}  \frac{
 \sum_{i=1}^{n}(X_{ij}-\bar{X}_j)^2  }
{  \sum_{i=1}^{n}(X_{ij}-\mu_{0j})^2 }
&= \prod_{j=1}^{p}\frac{(n-1)s_j^2}{(n-1)s_j^2 +n(\bar{X}_j-\mu_{0j})^2}\\
&={\prod_{j=1}^{p} \bigl\{1+\frac{n}{n-1}
(\bar{X}_j-\mu_{0j})^2  \bigr\}^{-1}}.
\end{align*}
This leads to the test statistic
\begin{equation*}
\begin{split}
-2\log(\Lambda_n )&=n\sum_{j=1}^{p}\mathrm{log}
\bigl\{1+\frac{n}{n-1}(\bar{X}_j-\mu_{0j})^2/s_j^2 \bigr\}=n\sum_{j=1}^{p}\log
\bigl(1+{t_{nj}^2}/{\nu_1} \bigr),
\end{split}
\end{equation*}
where $t_{nj}=\sqrt{n}(\bar{X}_j-{\mu}_{0j})/s_{j}$
are the standard $t$-statistics for the one-sample test
with $\nu_1=n-1$ degrees of freedom.

\vskip 12pt
\begin{lemma}\label{lamm0.1}
Let the gamma and digamma functions be the same as in Lemma 1.
For any $\nu>0$, we have the following integral equalities:
\begin{align*}
&\int_{0}^{\infty}(1+z^2)^{-(\nu+1)/2}dz=\frac{1}{2\sqrt{\nu} K(\nu)},\\
&\int_{0}^{\infty}(1+z^2)^{-(\nu+1)/2}\log(1+z^2)dz=\frac{D(\nu)}{2\sqrt{\nu} K(\nu)},\\
&\int_{0}^{\infty}(1+z^2)^{-(\nu+1)/2}\{\log(1+z^2)\}^2dz=
\frac{D^2(\nu)-2D'(\nu)}{2\sqrt{\nu} K(\nu)},
\end{align*}
where
$K(\nu)=\Gamma\{(\nu+1)/2\}/\{\sqrt{\pi \nu} \Gamma(\nu/2)\}$.
\end{lemma}

We note that this lemma essentially  follows  the results of Lemmas 4 to 6 in \citet{zhu2010}, and hence the proof is omitted.

\subsection{ Proof of Lemma 1} \label{appdixlemma1}
For simplicity, we omit the subscript $n$ in the terms  $U_{nj}$ and $t_{nj}$.
 Noting that $\nu_1=n-1$ and $U_j=n\log( 1+{t_{j}^2}/{\nu_1})$, where $t_{j}={\sqrt{n}(\bar{X}_j-\mu_{0j})}/{s_j}$, we have
\begin{align*}
E(U_j)&=\frac{2n}{\sqrt{\nu_1 \pi}}
 \frac{\Gamma\{(\nu_1+1)/2\}} {\Gamma(\nu_1/2)}\int_{0}^{\infty}
 \log\Big(1+\frac{t^2}{\nu_1}\Big) \Big(1+\frac{t^2}{\nu_1}\Big)^{-{\nu_1+1 \over 2}}dt,\\
E(U_j^2)&=\frac{2n^2}{\sqrt{\nu_1 \pi}}
\frac{\Gamma\{(\nu_1 +1)/2\}} {\Gamma(\nu_1 /2)}\int_{0}^{\infty}
\Bigl\{ \log\Big(1+\frac{t^2}{\nu_1}\Big) \Bigr\}^2  \Big(1+\frac{t^2}{\nu_1}\Big)^{-{\nu_1+1 \over 2}}dt.
 \end{align*}

\noindent Letting  $z=t/\sqrt{\nu_1}$ and by the method of substitution,
\begin{align*}
E(U_j)&=2n \sqrt{\nu_1} K(\nu_1) \int_{0}^{\infty} \log(1+z^2)(1+z^2)^{- \frac{\nu_1+1}{2} }dz,\\
E(U_j^2)&=2n^2 \sqrt{\nu_1} K(\nu_1)\int_{0}^{\infty} \big\{\log(1+z^2)\big\}^2(1+z^2)^{- \frac{\nu_1+1}{2} }dz.
\end{align*}
By  Lemma \ref{lamm0.1}, we have
$E(U_j)=nD(\nu_1)$ and $E(U_j^2)=n^2\big\{D^2(\nu_1)-2D'(\nu_1)\big\}.$
This shows that
$E(U_j)= m_1$ and $E(U_j^2) = m_2$, consequently, $\mathrm{Var}(U_j)= m_2-m_1^2.$

We note that $n\log(1+{t^2}/{\nu_1})\leq {n t^2}/{\nu_1}<2t^2$,
and that $(1+{t^2}/{\nu_1})^{-n/2}$ converges to
$e^{-{t^2}/{2}}$ as ${n \to \infty}$.
By the dominated convergence theorem,
\begin{equation*}\label{m1.rate}
\begin{split}
&\lim_{n \to \infty} \frac{2n}{\sqrt{\nu_1 \pi}}
 \frac{\Gamma\{(\nu_1+1)/2\}} {\Gamma(\nu_1/2)}\int_{0}^{\infty}
 \log\Big(1+\frac{t^2}{\nu_1}\Big) \Big(1+\frac{t^2}{\nu_1}\Big)^{-\frac{\nu_1+1}{2}}dt\\
&=   \lim_{n \to \infty} \frac{2}{\sqrt{\nu_1 \pi}}
 \frac{\Gamma\big\{(\nu_1 +1)/2\big\}} {\Gamma(\nu_1/2)}
 \int_{0}^{\infty} t^2e^{-\frac{t^2}{2}}dt\\
&= \sqrt{\frac{2}{\nu_1}}  \lim_{n \to \infty}
 \frac{\Gamma\big\{(\nu_1+1)/2\big\}} {\Gamma(\nu_1/2)}=1,
\end{split}
\end{equation*}
where the last equation is obtained by Stirling's formula,
$\Gamma(x)=\sqrt{2\pi}x^{x-1/2}e^{-x}\{1+O(1/x)\}$ as $x\to\infty$ \citep{spira1971}.
  This shows that $m_1 \to 1$ as $n \to \infty$.
Similarly, $m_2 \to 3$ as $n \to \infty$.
Finally, $\mathrm{Var}(U_j)= m_2-m_1^2 \to 2$ as $n \to \infty$.

\subsection{Proof of Theorem 1}\label{AppendixTh1}

As the sequence $\{ U_{nj} \}$ satisfies conditions (C1) and (C2), we only need to prove that $E|U_{nj}-E(U_{nj})|^{2+\delta}< \infty$ for any fixed $n$. We note that
\begin{align*}
E|U_{nj}-E(U_{nj})|^{2+\delta}
={2n^{2+\delta} \over \sqrt{\nu_1 \pi}}
 \frac{\Gamma\{(\nu_1+1)/2\}} {\Gamma(\nu_1/2)} \int_{0}^{\infty}\Big(1+{t^2\over \nu_1}\Big)^{-{\nu_1+1 \over 2}}\Big\{\log{\Big(1+{t^2\over \nu_1}\Big)}-D(\nu_1)\Big\}^{2+\delta} dt.
\end{align*}
Then we only need to verify that,  for any fixed $n\geq2$,
\begin{align*}
\int_{0}^{\infty}(1+x^2)^{-{\nu_1+1 \over 2}}\{\log{(1+x^2)}-D(\nu_1)\}^{2+\delta} dx<\infty.
\end{align*}
The inequality clearly holds.

Finally, by the central limit theorem under the strong mixing condition  (see Corollary 5.1 in  \citet{hall2014}),  we have
${({T}_1-p m_1)}/({\tau_{1}\sqrt{p}})
{\overset{\mathcal{D} }\longrightarrow} N(0,1)$ as ${p \to \infty}$.

\subsection{Proof of Corollary 1}\label{AppendixTh2}

(a) When the covariance matrix is a diagonal matrix,
$\{t_{nj}={\sqrt{n}(\bar{X}_j-\mu_{0j})}/{s_j},~j=1,\ldots,p \}$ are i.i.d. random variables. Consequently, $\{U_{nj}, ~j=1,\ldots, p \}$ are also i.i.d. random variables with mean $m_1$ and variance $m_2-m_1^2$.
By the central limit theorem, we have

\noindent
$({\widetilde{T}_1-m_1})/\sqrt{p(m_2-m_1^2)}
{~\overset{\mathcal{D}}\longrightarrow~} N(0,1)$ as $p \to \infty$.
\vskip 12pt
\noindent(b) For simplicity, we omit the subscript $n$  in the terms $U_{nj}$ and $t_{nj}$. We note that
$\log(1+t_j^2/\nu_1)$ is sandwiched between
${t_j^2}/{\nu_1}  - {(t_j^2)^2}/{2\nu_1^2} +\cdots+(-1)^{k+2}{( t_j^2)^{k+1}}/{\{(k+1)\nu_1^{k+1}\}}$
and ${t_j^2}/{\nu_1} -{(t_j^2)^2}/{2\nu_1^2} +\cdots+(-1)^{k+3}{( t_j^2)^{k+2}}/{\{(k+2)\nu_1^{k+2}\}}$. For any $n>2k+6$,
\begin{align*}
%&E (U_j)  \geq \frac{n}{\nu_1} E(t_j^2) - \frac{n}{2\nu_1^2}E\{( %t_j^2)^2\}+\cdots-\frac{n}{(k+2)\nu_1^{k+2} }E\{( t_j^2)^{k+2}\}\\
%&E (U_j)  \leq \frac{n}{\nu_1} E(t_j^2) - \frac{n}{2 \nu_1^2}E\{( %t_j^2)^2\}-\cdots+\frac{n}{(k+1)\nu_1 ^{k+1} }E\{( t_j^2)^{k+1}\}\\
&E (U_j)= m_1 =\frac{n}{\nu_1} E(t_j^2) - \frac{n}{2\nu_1^2}E\{( t_j^2)^2\}
  +\cdots+(-1)^{k+1}\frac{n}{k\nu_1^{k} }E\{( t_j^2)^{k}\}+O(1/n^{k}). \nonumber
\end{align*}
Noting that $E\{( t_j^2/\nu_1)^{k}\}=a_k$, and $m_2-m_1\to2$ as $n\to \infty$,  we have
\begin{align*}
\frac{{T}_1-p \xi_k}{\sqrt{2p}}
&=\frac{\sqrt{p}(\widetilde{T}_1-m_1)}{\sqrt{2}}
+\frac{\sqrt{p}(m_1-\xi_k)}{\sqrt{2}}\\
&=\frac{\sqrt{p}(\widetilde{T}_1-m_1)}
{\sqrt{m_2-m_1^2}}
\frac{\sqrt{m_2-m_1^2}}{\sqrt{2}}+\sqrt{p}O(1/n^k)
~{\overset{\mathcal{D}}\longrightarrow}~N(0,1).
\end{align*}

\subsection{Proof of Theorem 2}\label{Appendixpower1}

First of all, we have
\begin{equation}\label{center.H1}
\begin{split}
\frac{\widetilde{T}_1- 1}
{\sqrt{\hat{\tau}_{1}^2}}
&=\frac{
\widetilde{T}_1-\overline{m}_{1}}{\sqrt{\hat{\tau}_{1}^2}}+ \frac{\sqrt{p}(\overline{m}_{1}-1)}{\sqrt{\hat{\tau}_{1}^2}},
\end{split}
\end{equation}
where
$\overline{m}_{1}=\sum_{j=1}^{p} \widetilde{m}_{1j}/p$ and
$\widetilde{m}_{1j}=E(U_{j}| H_1)=E\big\{n\mathrm{log}\big(1+
t_j^2/\nu_1\big)| H_1 \big\}$.

We note that
${n} t_j^2/{\nu_1} - {n}( t_j^2)^2/{(2\nu_1^2)}
\leq U_j \leq {n} t_j^2/{\nu_1}$. Then, for any $n>6$, we have
${n} E(t_j^2| H_1)/{\nu_1} - {n}E\big\{( t_j^2)^2  | H_1\big\}/{(2\nu_1^2)}
\leq E (U_j | H_1)   \leq  {n}E( t_j^2  | H_1)/ {\nu_1}$.
Under the local alternative (5) and  condition (6),  we have $\widetilde{m}_{1j}={n} E(t_j^2 | H_1)/{\nu_1} +O(1/n)$.
We also note that $t_{j}$ follows a noncentral $t$ distribution with  $\nu_1=n-1$ degrees of freedom and a noncentrality  parameter $\Delta_{1j}$. Its second moment is $E(t_{j}^2| H_1)=\nu_1(1+\Delta_{1j}^2)/(\nu_1-2)$; hence, $\widetilde{m}_{1j}=1+\Delta_{1j}^2 +O(1/n)$. Consequently,
$\overline{m}_1=\sum_{j=1}^{p}\widetilde{m}_{1j}/p=1+\bm{\Delta}_1^T \bm{\Delta}_1/{p}+O(1/n)$.

Under conditions (C1) and (C2), for any consistent estimator $\hat{\tau}_{1}^2$ for $\tau_{1}^2$,
\begin{align*}
\frac{{T}_1-p}
{\sqrt{p \hat{\tau}_{1}^2}}
&=\frac{\sqrt{p}
(\widetilde{T}_1-\overline{m}_{1})}{\sqrt{\hat{\tau}_{1}^2}}
+\frac{\bm{\Delta}_1^T \bm{\Delta}_1/\sqrt{p}}{{\sqrt{\hat{\tau}_{1}^2}}} +\frac{\sqrt{p}O(1/n)}{\sqrt{\hat{\tau}_{1}^2}}\\
%&=\frac{\sqrt{p}(
%\widetilde{T}_1-E(\widetilde{T}_1 |H_1))}
%{\sqrt{\hat{\tau}_{1}^2}}
%+\frac{\bm{\Delta}_1^T \bm{\Delta}_1/\sqrt{p} %+\sqrt{p}O(1/n)}{\sqrt{\hat{\tau}_{1}^2}}\\
&{\overset{\mathcal{D}}\longrightarrow}~N(0,1)+\frac{\bm{\Delta}_1^T \bm{\Delta}_1/\sqrt{p}}{\sqrt{{\tau}_{1}^2}}.
\end{align*}
We note that $\bm{\Delta}_1^T\bm{\Delta}_1/\sqrt{p}
=\sum_{j=1}^{p}(\delta_{1j}/\sigma_{jj})^2 /\sqrt{p}$.
Thus, if $\sqrt{p}=o(\sum_{j=1}^{p}\delta_{1j}^2/\sigma_{jj}^2)$,
then the power of the one-sample test will increase towards 1 as $(n, p)\to\infty$. On the other hand, if $\sum_{j=1}^{p}\delta_{1j}^2/\sigma_{jj}^2=o(\sqrt{p})$,
then the test will have little power as $(n,p)\to \infty$.

\renewcommand\thesection{{B}}{}
\section{DLRT for Two-Sample Test}
\renewcommand\thesection{{B}}{}
\subsection{Derivation of Formula (8)}\label{stat2}
For the two-sample test,
we derive the DLRT statistic based on the assumption that
the two covariance matrices are equal and they follow a diagonal matrix structure, i.e., $\Sigma_1=\Sigma_2= \Sigma={\text{diag}}(\sigma_{11}^2,\sigma_{22}^2,\ldots, \sigma_{pp}^2)$.

Under the alternative hypothesis,  the maximum of the likelihood function is
\begin{align*}
~&\underset{\bm{\mu}_1,\bm{\mu}_2,\Sigma}{\mathrm{max}}L(\bm{\mu}_1,
\bm{\mu}_2, \Sigma|H_1)\\
&=(2 \pi e)^{{-{p\over 2}(n_1+n_2)}}
\prod_{j=1}^{p}
\Bigl[\frac{1}{n_1+n_2}\Bigl\{
\sum_{i=1}^{n_1}(X_{ij}-\bar{X}_{j})^2 +
\sum_{k=1}^{n_2}(Y_{kj}-\bar{Y}_{j})^2 \Bigl\} \Bigr]^{-{(n_1+n_2)\over 2}}.
\end{align*}
Similarly, under the null hypothesis $H_0: \bm{\mu}_1=\bm{\mu}_2$,
\begin{align*}
~&\underset{\bm{\mu}_1,\bm{\mu}_2,\Sigma}{\mathrm{max}}L(\bm{\mu}_1,
\bm{\mu}_2, \Sigma| H_0)\\
&=(2 \pi e)^{-{p\over2}(n_1+n_2)} \prod_{j=1}^{p}
\Bigl[\frac{1}{n_1+n_2}\Bigl\{
\sum_{i=1}^{n_1}(X_{ij}-\widehat{\mu}_{0j})^2 +
\sum_{k=1}^{n_2}(Y_{kj}-\widehat{\mu}_{0j})^2 \Bigl\} \Bigr]^{-{(n_1+n_2)\over 2}},
\end{align*}
where $\widehat{\mu}_{0j}=(n_1\bar{X}_j+n_2\bar{Y}_j)/(n_1+n_2)$,
$\bar{X}_j=\sum_{i=1}^{n_1}X_{ij}/n_1$, and
$\bar{Y}_j=\sum_{k=1}^{n_2}Y_{kj}/n_2$.

We also note that
\begin{align*}
~&\sum_{i=1}^{n_1}(X_{ij}-\widehat{\mu}_{0j})^2 +
\sum_{k=1}^{n_2}(Y_{kj}-\widehat{\mu}_{0j})^2\\
&=\sum_{i=1}^{n_1}(X_{ij}-\bar{X}_{j})^2+
\sum_{k=1}^{n_2}(Y_{kj}-\bar{Y}_{j})^2
+\frac{n_1n_2^2}{(n_1+n_2)^2}(\bar{X}_{j}-\bar{Y}_{j})^2
+\frac{n_2n_1^2}{(n_1+n_2)^2}(\bar{Y}_{j}-\bar{X}_{j})^2.
\end{align*}
We have
\begin{align*}
\Lambda_N&=\frac{\underset{\bm{\mu_1},\bm{\mu_2},\Sigma}{\mathrm{max}} L(\bm{\mu_1},\bm{\mu_2},\Sigma|H_0)}
{\underset{\bm{\mu_1},\bm{\mu_2},\Sigma}{\mathrm{max}} L(\bm{\mu_1},\bm{\mu_2},\Sigma|H_1)}
=\prod_{j=1}^{p}\frac{
\bigl\{ \frac{n_1+n_2-2}{n_1+n_2} s_{j,\mathrm{pool}}^2 +
\frac{n_1 n_2}{(n_1+n_2)^2} (\bar{X}_j-\bar{Y}_j)^2 \bigr\}^{-{(n_1+n_2)\over 2}}}
{\Bigl[  \frac{1}{n_1+n_2}\big\{
\sum_{i=1}^{n_1}(X_{ij}-\bar{X}_{j})^2 +
\sum_{k=1}^{n_2}(Y_{kj}-\bar{Y}_{j})^2 \big\} \Bigr]^{-{(n_1+n_2)\over 2}}  },
\end{align*}
where
$s_{j,\mathrm{pool}}^2=
\{(n_1-1)s_{1j}^2+(n_2-1)s_{2j}^2\}/{(n_1+n_2-2)}$
are the pooled sample variances with
$s_{1j}^2=\sum_{i=1}^{n_1}(X_{ij}-\bar{X}_j)^2/(n_1-1)$
and $s_{2j}^2=\sum_{k=1}^{n_2}(Y_{kj}-\bar{Y}_j)^2/(n_2-1)$.
 This leads to the test statistic
\begin{align*}
-2\log(\Lambda_N)
&=(n_1+n_2)\sum_{j=1}^{p}{\log}
\Bigl\{1+
\frac{\frac{n_1n_2}{n_1+n_2}(\bar{X}_j-\bar{Y}_{j})^2}{s_{j,\mathrm{pool}}^2}
\frac{1}{n_1+n_2-2} \Bigr\}=N\sum_{j=1}^{p}{\log}
\Bigl(1+\frac{t_{Nj}^2}{\nu_2}\Bigr),
\end{align*}
where $t_{Nj}=\sqrt{n_1n_2/(n_1+n_2)}(\bar{X}_j-\bar{Y}_{j})/s_{j,\mathrm{pool}}$
are the standard $t$-statistics for the two-sample test with $\nu_2=N-2$ degrees of freedom.

\subsection{ Proof of  Theorem 3}\label{AppendixTh4}

We first show that
$E(V_{Nj})= G_1 \to 1$ and
$\mathrm{Var}(V_{Nj}) =  G_2-G_1^2 \to 2$ as $N \to \infty.$
For simplicity, we omit the subscript $N$  in the terms $V_{Nj}$ and $t_{Nj}$.
 As $V_j=N\log\bigl( 1+{t_{j}^2}/{\nu_2} \bigr)$, we have
\begin{align*}
E(V_{j})&=\frac{2N}{\sqrt{\nu_2 \pi}}
 \frac{\Gamma\{(\nu_2+1)/2\}} {\Gamma(\nu_2/2)}\int_{0}^{\infty}
 \log\Big(1+\frac{t^2}{\nu_2}\Big) \Big(1+\frac{t^2}{\nu_2}\Big)^{-{\nu_2+1 \over 2}}dt,\\
E(V_{j}^2)&=\frac{2N^2}{\sqrt{\nu_2 \pi}}
 \frac{\Gamma\{(\nu_2 +1)/2\}} {\Gamma(\nu_2 /2)}\int_{0}^{\infty}
 \Bigl\{ \log\Big(1+\frac{t^2}{\nu_2}\Big) \Bigr\}^2  \Bigl(1+\frac{t^2}{\nu_2}\Bigr)^{-{\nu_2+1 \over 2}}dt.
\end{align*}
Let $z=t/\sqrt{\nu_2}$, we have
\begin{align*}
E(V_{j})&=2N \sqrt{\nu_2} K(\nu_2) \int_{0}^{\infty} \log(1+z^2)(1+z^2)^{- \frac{\nu_2+1}{2} }dz,\\
E(V_{j}^2)&=2N^2 \sqrt{\nu_2} K(\nu_2)\int_{0}^{\infty} \big\{\log(1+z^2)\big\}^2(1+z^2)^{- \frac{\nu_2+1}{2}}dz.
\end{align*}
By Lemma \ref{lamm0.1}, we have
$E(V_{j})= G_1=ND(\nu_2)$ and $E(V_{j}^2)= G_2=N^2\{D^2(\nu_2)-2D'(\nu_2)\}$.

We note that $N\log(1+{t^2}/{\nu_2})\leq{N t^2}/{\nu_2}\leq 2t^2$,
and that $(1+{t^2}/{\nu_2})^{-N/2}$  converges to $e^{-{t^2}/{2}}$
as $N \to \infty$.
By the dominated convergence theorem,
\begin{equation*}\label{m1.rate}
\begin{split}
&\lim_{N \to \infty} \frac{2N}{\sqrt{\nu_2 \pi}}
 \frac{\Gamma\{(\nu_2+1)/2\}} {\Gamma(\nu_2/2)}\int_{0}^{\infty}
 \log\Big(1+\frac{t^2}{\nu_2}\Big) \Big(1+\frac{t^2}{\nu_2}\Big)^{-\frac{\nu_1+1}{2}}dt\\
&=   \lim_{N \to \infty} \frac{2}{\sqrt{\nu_2 \pi}}
 \frac{\Gamma\{(\nu_2 +1)/2\}} {\Gamma(\nu_2/2)}
 \int_{0}^{\infty} t^2e^{-\frac{t^2}{2}}dt\\
&=\sqrt{\frac{{2}}{\nu_2}} \lim_{N \to \infty}
 \frac{\Gamma\{(\nu_2+1)/2\}} {\Gamma(\nu_2/2)}=1.
\end{split}
\end{equation*}
This shows that $E(V_{j})= G_1\to 1$ as $N\to\infty$. Similarly, $E(V_{j}^2)= G_2 \to 3$ as $N\to\infty$.
Finally, $\mathrm{Var}(V_{j})= G_2-G_1^2 \to 2$ as $N\to\infty$.

\vskip12pt
 Second, we prove that
 $({T}_2-p G_1)/{(\tau_{2}\sqrt{p})}{~\overset{\mathcal{D}}\longrightarrow~} N(0,1)$
as ${p\rightarrow\infty}$.
As the sequence $\{ V_{j} \}$ satisfies conditions (C1) and (C2), we only need to prove that, for any fixed $N \geq 4$,   $E|V_{j}-E(V_{j})|^{2+\delta}< \infty$.
We note that
\begin{align*}
E|V_{j}-E(V_{j})|^{2+\delta}
={2N^{2+\delta} \over \sqrt{\nu_2 \pi}}
 \frac{\Gamma\{(\nu_2+1)/2\}} {\Gamma(\nu_2/2)} \int_{0}^{\infty}\Big(1+{t^2\over \nu_2}\Big)^{-{\nu_2+1 \over 2}}\Big\{\log{\Big(1+{t^2\over \nu_2}\Big)}-D(\nu_2)\Big\}^{2+\delta} dt.
\end{align*}
Then, by the similar arguments to those in the proof of Theorem 1, we have
$E|V_{j}-E(V_{j})|^{2+\delta}<\infty$.

\subsection{ Proof of  Corollary 2}\label{AppendixTh5}
\noindent(a) When the covariance matrix is a diagonal matrix, $\{ t_{Nj}=\sqrt{n_1n_2/(n_1+n_2)}(\bar{X}_j-\bar{Y}_{j})/s_{j,\mathrm{pool}},
 j=1,\ldots,p \}$ are i.i.d. random variables. Consequently, $\{ V_{Nj}, j=1,\ldots, p \}$ are also i.i.d. random variables with mean $G_1$ and variance $G_2-G_1^2$. By the central limit theorem, we have
$({{T}_2-pG_1})/\sqrt{p(G_2-G_1^2)}
~{\overset{\mathcal{D}}\longrightarrow}~ N(0,1)$ as ${p\to\infty}$.

\vskip 12pt
\noindent(b)
For simplicity, we omit the subscript $N$  in the terms $t_{Nj}$ and $V_{Nj}$.
We note that
$\log(1+t_j^2/\nu_2)$ is sandwiched between
${t_j^2}/{\nu_2}  - {(t_j^2)^2}/{2\nu_2^2} +\cdots+(-1)^{k+2}{( t_j^2)^{k+1}}/{\{(k+1)\nu_2^{k+1}\}}$
and ${t_j^2}/{\nu_2} -{(t_j^2)^2}/{2\nu_2^2} +\cdots+(-1)^{k+3}{( t_j^2)^{k+2}}/{\{(k+2)\nu_2^{k+2}\}}$. Then, for any $N>2k+6$,
\begin{align}
&E (V_j) = G_1=\frac{N}{\nu_2} E(t_j^2) - \frac{N}{2\nu_2^2}E\{( t_j^2)^2\}-\cdots-\frac{N}{k\nu_2^{k} }E\{( t_j^2)^{k}\}+O(1/N^{k}). \nonumber
\end{align}
Following the proof of Theorem 3, we have
$G_2-G_1^2 \to 2$ as ${N\to\infty}$. Further,
\begin{align*}
\frac{{T}_2-p \xi_k}{\sqrt{2p}}
&=\frac{\sqrt{p}({T}_2/p-G_1)}{\sqrt{2}}
+\frac{\sqrt{p}(G_1-\eta_k)}{\sqrt{2}}\\
&=\frac{\sqrt{p}({T}_2/p-G_1)}
{\sqrt{G_2-G_1^2}}\frac{\sqrt{G_2-G_1^2}}{\sqrt{2}}+\sqrt{p}O(1/N^k)
~{\overset{\mathcal{D}}\longrightarrow}~N(0,1).
\end{align*}

\subsection{ Proof of  Theorem 4}\label{Appendixpower2}
First of all, we have
\begin{align}\label{center.H2}
\frac{\widetilde{T}_2- 1}
{\sqrt{\hat{\tau}_{2}^2}}
=\frac{
\widetilde{T}_2-\overline{G}_{1}}{\sqrt{\hat{\tau}_{2}^2}}+ \frac{\overline{G}_{1}-1}{\sqrt{\hat{\tau}_{2}^2}},
\end{align}
where $\overline{G}_{1}=\sum_{j=1}^{p} \widetilde{G}_{1j}/p$
and $\widetilde{G}_{1j}=E(V_{j}| H_1)=
E\{N\log\bigl(1+ t_j^2/\nu_2\bigr)|H_1\}$.

We note that  ${N}\{t_j^2/{\nu_2}-( t_j^2)^2/({2\nu_2^2})\}
\leq V_j \leq {N}t_j^2/{\nu_2}$.  Then, for any $N>6$, we have
${N} [ E(t_j^2|H_1)/{\nu_2}-E\{(t_j^2)^2 | H_1\}/({2\nu_2^2})]\leq E (V_j | H_1)\leq {N}E(t_j^2| H_1)/{\nu_2}$.
Under the local alternative (9) and condition (10), we have $\widetilde{G}_{1j}= {N} E(t_j^2 | H_1)/{\nu_2} +O(1/N)$.
We also note that $t_{j}$ follows a noncentral $t$ distribution with $\nu_2=N-2$ degrees of freedom and a noncentrality parameter $\Delta_{2j}$. Its second moment is $E(t_{j}^2| H_1)=\nu_2(1+\Delta_{2j}^2)/(\nu_2-2)$;   hence, $\widetilde{G}_{1j}=1+\Delta_{2j}^2 +O(1/N)$. Consequently, $\overline{G}_{1}=1+{\bm{\Delta}_2^T \bm{\Delta}_2/p} +O(1/N)$.

Under conditions (C1) and (C2), for any consistent estimator $\hat{\tau}_{2}^2$ for $\tau_{2}^2$,
\begin{align*}
\frac{{T}_2-p}
{\sqrt{p \hat{\tau}_{2}^2}}
&=\frac{\sqrt{p}(
\widetilde{T}_2-\overline{G}_{1})}{\sqrt{\hat{\tau}_{2}^2}}
+\frac{\bm{\Delta}_2^T \bm{\Delta}_2/\sqrt{p}}{{\sqrt{\hat{\tau}_{2}^2}}}
+\frac{\sqrt{p}O(1/N)}{\sqrt{\hat{\tau}_{2}^2}}\\
&{\overset{\mathcal{D}}\longrightarrow}~
 N(0,1)+\frac{\bm{\Delta}_2^T \bm{\Delta}_2/\sqrt{p}}{\sqrt{{\tau}_{2}^2}}.
\end{align*}
We note that $\bm{\Delta}_2^T\bm{\Delta}_2/\sqrt{p}
=p^{-1/2}\sum_{j=1}^{p}(\delta_{2j}^2/\sigma_{jj}^2)$.
Thus, if $\sqrt{p}=o(\sum_{j=1}^{p}\delta_{2j}^2/\sigma_{jj}^2)$,
then the power of the two-sample test will increase  towards 1 as $(N, p)\to \infty$. On the other hand, if $\sum_{j=1}^{p}\delta_{2j}^2/\sigma_{jj}^2=o(\sqrt{p})$,
then the test will have little power as $(N,p)\to \infty$.
\end{appendices}

\renewcommand\thesection{{C}}{}
\section{Additional Simulations}
\subsection{Simulations for the comparison between the DLRT and PA tests}

As  mentioned by the reviewer,  \citet{park2013test} also proposed a scale invariant test, referred to as the PA test, which uses the idea of leave-out cross validation.
To evaluate the performance of the DLRT and PA tests, we conduct simulations based on the same settings as Section 4.1 in the main paper.

\begin{figure}[!htbp]
    \centering
    \subfigure[$n_1=n_2=5, p=100$]
    {\includegraphics[width=16cm,height=9cm]{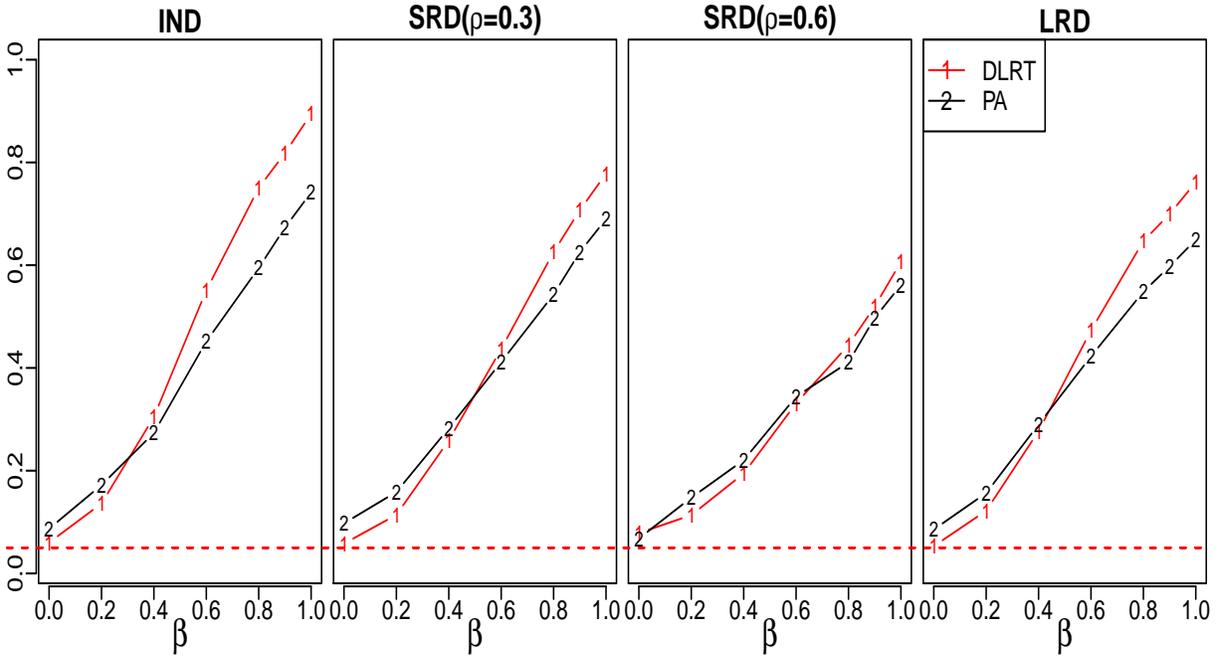}}\\
    \subfigure[$n_1=n_2=15, p=500$]
    {\includegraphics[width=16cm,height=9cm]{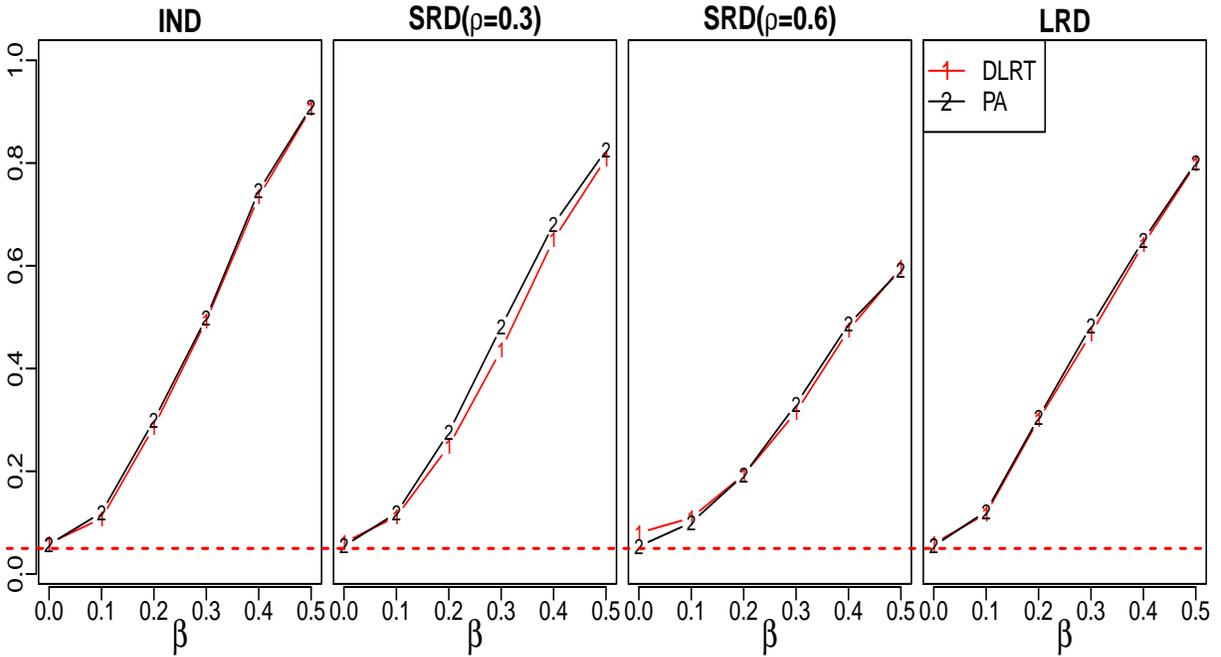}}
    \caption{\label{PApower.n=5} Power comparisons between the DLRT and PA tests with ($n_1=n_2=5, p=100$) or ($n_1=n_2=15, p=500$), respectively.  The horizontal dashed red lines represent the nominal level of $\alpha=0.05$. The results are based on 2000 simulations with data from the normal distribution.}
\end{figure}

Figure \ref{PApower.n=5} presents the  power of the DLRT and PA tests with the nominal level of $\alpha=0.05$. When the sample size is small (e.g., $n=5$), the PA test has some inflated type I error rate,
and also suffers from a low power of detection.
When the sample size is not small (e.g., $n=15$), the DLRT and PA tests are able to simultaneously control the type I error rate, and at the same time keep a high power of detection.
To conclude, when the sample size is not large, DLRT performs better than or at least as well as the PA test for normal distributed data.

\linespread{1}

\linespread{1.25}

In addition, the PA test is not a shift invariant test, which indicates that if the mean vectors under the null hypothesis are not located at the origin,  the PA test may not keep the type I error rate at the nominal level.
To further demonstrate this point, let $\bm{\mu}_1=\bm{\mu}_2=(100,\cdots, 100)^T\in R^{p}$.
We consider $p=100$, $300$ or $500$ and $n_1=n_2=5$ or $15$, respectively.
The other settings are the same as Section 4.1 in the main paper.
The simulation results are reported in Table \ref{tabShift}.
From Table \ref{tabShift},  we can see that the PA test suffers from significant inflated type I error rate,
whereas, the DLRT test still has a well control type I error rate.

\linespread{1}

\begin{center}
{
\begin{table}[htbp!]
\caption{\label{tabShift}  Type I error rates over 2000 simulations for the DLRT and PA tests under the IND or SRD structure. The nominal level is $\alpha=0.05$. In the SRD structure, we consider the correlation, $\rho=0.6$.}
{  {
\begin{center}
\begin{tabular}{c|c||c|c|c||c|c|c}
\hline\hline
& & \multicolumn{3}{c||}{$n_1=n_2=5$} & \multicolumn{3}{c}{$n_1=n_2=15$} \\\hline
\multirow{2}{*}{Cov}& \multirow{2}{*}{Method} & \multirow{2}{*}{$p=100$}
& \multirow{2}{*}{$p=300$}& \multirow{2}{*}{$p=500$}&
\multirow{2}{*}{$p=100$} & \multirow{2}{*}{$p=300$}& \multirow{2}{*}{$p=500$}\\
&  & & & & & & \\\hline
\multirow{2}{*}{IND}
&DLRT&0.056&0.052&0.043&0.058&0.049 & 0.048 \\\cline{2-8}
&PA&0.981&0.985&0.987&0.763&0.795 & 0.779  \\ \hline\hline
\multirow{1}{*}{SRD}
&DLRT&0.076&0.080&0.072&0.078&0.081 &0.078    \\\cline{2-8}
$\rho=0.6$
&PA& 0.969  &0.980  & 0.967 & 0.601 & 0.649 &0.631   \\ \hline\hline
\end{tabular}\end{center}
} }
\end{table}
}
\end{center}

\linespread{1.25}

\subsection{Additional simulations with highly heterogeneous variances}

In Section 4.1 in the main paper,  the variances, $\sigma_{11}^2,\ldots,\sigma_{pp}^2$, are randomly sampled from the scaled chi-square distribution $\chi_{5}^2/5$. To  account for highly heterogeneous variances,
we let $\sigma_{11}^2<\sigma_{11}^2<\ldots<\sigma_{pp}^2$ be equally spaced on the interval $[0.01,150]$.
The other settings are all the same as  Section 4.1.
Accordingly, the variances of the signal components are small, whereas the variances of the noise components are large.

\begin{figure}[!htbp]
    \centering
    \subfigure[$n_1=n_2=5, p=100$]
    {\includegraphics[width=16cm,height=9.5cm]{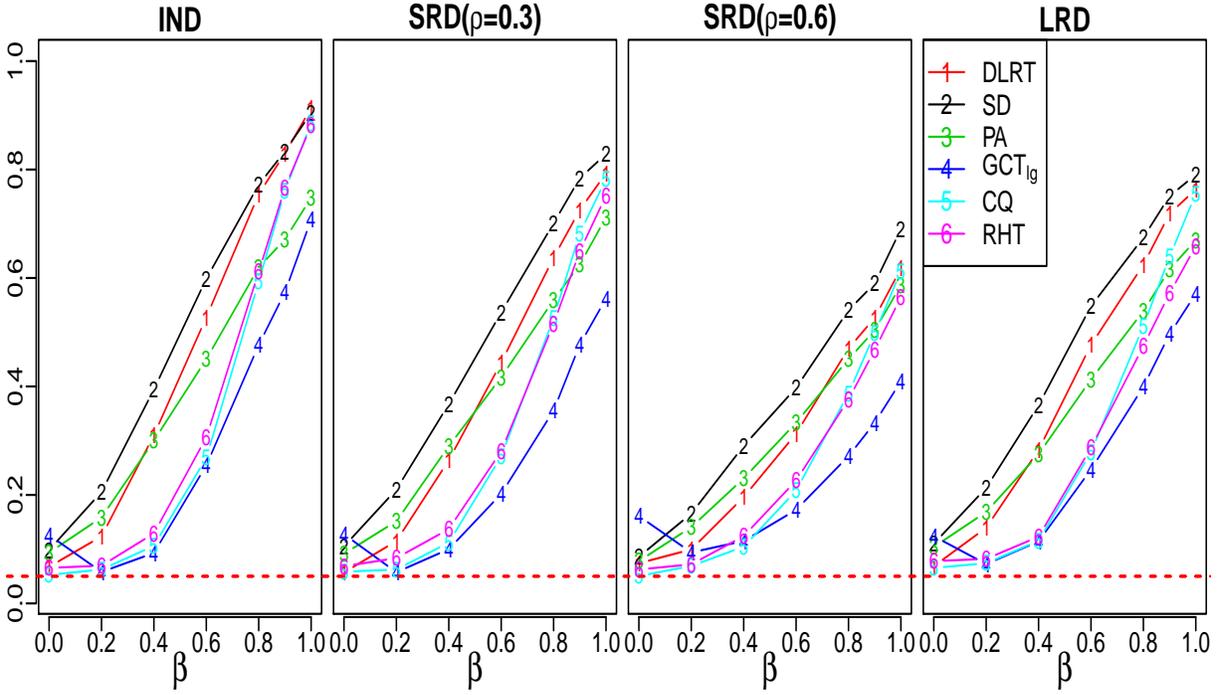}}\\
    \subfigure[$n_1=n_2=15, p=500$]
    {\includegraphics[width=16cm,height=9.5cm]{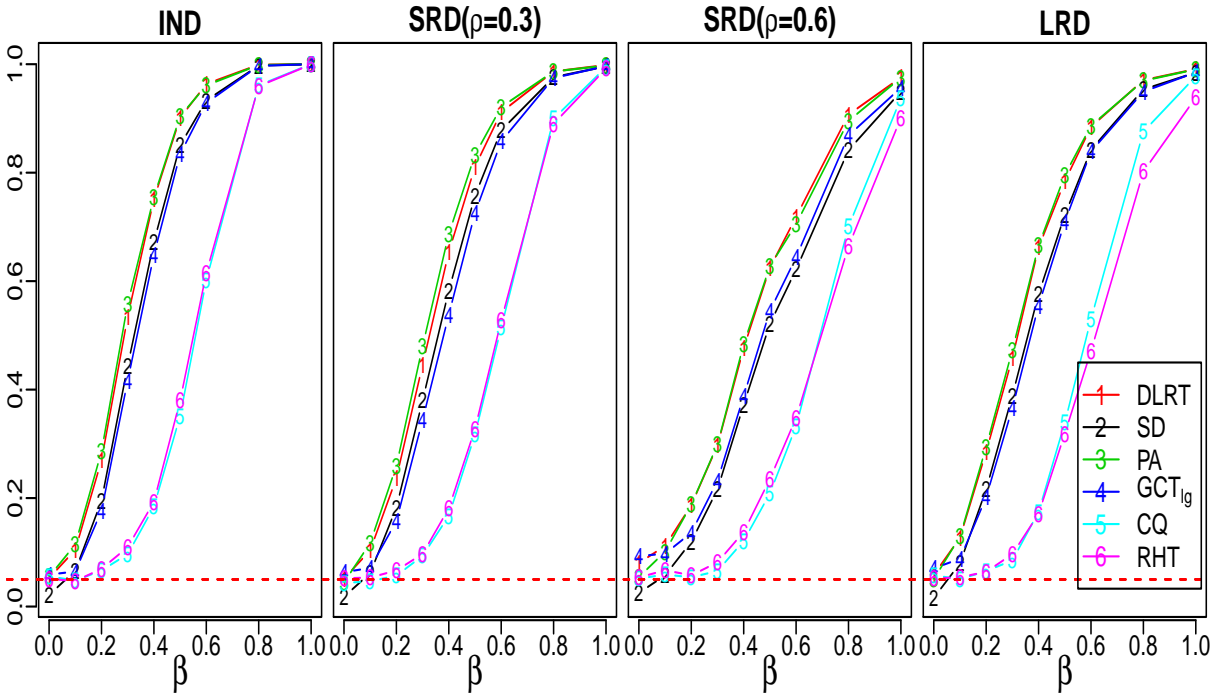}}
    \caption{\label{Figvar1} Power comparisons among the DLRT, SD, PA, ${\rm GCT}_{\rm lg}$, CQ and RHT tests with ($n_1=n_2=5, p=100$) or ($n_1=n_2=15, p=500$), respectively.
    The horizontal dashed red lines represent the nominal level of $\alpha=0.05$. The results are based on 2000 simulations with data from the normal distribution. }
\end{figure}

Figure \ref{Figvar1} shows the simulation results. When the sample size is small, e.g., $n=5$, the SD, PA, ${\rm GCT}_{\rm lg}$ tests suffer from inflated type I error rates, whereas, the DLRT is able to control the type I error rate and exhibits high power than the CQ and RHT tests. As the dimension is large and the sample size is not small, the DLRT and PA tests are both able to control the type I error rates and at the same time exhibit high power than the other tests such as CQ and RHT. In addition, from Figure \ref{Figvar1}, we can also see that the diagonal Hotelling's tests like our proposed DLRT perform better than the unscaled  Hotelling's tests like CQ, and the regularized  Hotelling's tests like RHT. As mentioned in the Introduction of the main paper,  when the variances of the signal components are small and the variances of the noise components are large, the scale transformation invariant tests such as the diagonal Hotelling's tests usually provide a better performance than the orthogonal transformation invariant tests such as the unscaled Hotelling's tests and the regularized Hotelling's tests.

\linespread{1}

\linespread{1.25}

\subsection{Additional simulations with general multivariate data}

The proposed  new test assumes a natural ordering of the components in the $p$-dimensional random vector, e.g., the correlation among the components are related to their positions, and hence we can take into account the additional structure information to avoid an estimation of the full covariance matrix.
When the ordering of the components is not available, we propose to reorder the components from the sample data using some well known ordering methods before applying our proposed tests. For instance, with the best permutation algorithm in \citet{rajaratnam2013}, the strongly correlated elements can be reordered
close to each other. For other  ordering methods of  random variables, one may refer to, for example, \citet{gilbert1992}, \citet{wagaman2009}, and the references therein.

~~We have also conducted simulations to evaluate the performance of DLRT after reordering the components by the best permutation algorithm. Specifically, we first generate $\bm{X}_1,\ldots,\bm{X}_{n_1}$ from $N_{p}(\bm{\mu}_1,\Sigma)$, and $\bm{Y}_1,\ldots,\bm{Y}_{n_2}$ from $N_{p}(\bm{\mu}_2,\Sigma)$,
where the common covariance matrix, $\Sigma$, follows the same setting as Section 4.1 of our main paper.
For simplicity, let $\bm{\mu}_1=\bm{0}$. Under the alternative hypothesis, we assume that the first $p_0$ elements in $\bm{\mu}_2$ are nonzero, where $p_0=\beta p$ with  $\beta \in [0,1]$ being the tuning parameter that controls  the signal sparsity. We consider two dependence structures including the independent (IND) structure and the short range dependence (SRD) structure. To represent different levels of correlation in the SRD structure, we consider two levels of correlation $\rho=0.3$ and $0.6$, respectively.
For the power comparison, we set the $j$th nonzero component in $\bm{\mu}_2$ as $\mu_{2j}=\theta \sigma_{jj}, j=1,\ldots, p_0$, where $\theta$ is the effect size of the corresponding component.
The other parameters are set as  $(n_1,n_2,\theta)  \times  p = \{(30,30,0.25)\} \times  \{500 {~\rm or~} 800\}$, respectively.
In addition, we let $\widetilde{X}_i=P\bm{X}_i$ and $\widetilde{Y}_k=P\bm{Y}_k$, where $P\in R^{p\times p}$ is a permutation matrix.  Consequently, the components of $\widetilde{X}_i$ and $\widetilde{Y}_k$ do not follow a natural ordering.

Next, we apply the BP-DLRT,  SD, PA, BP-${\rm GCT}_{\rm lg}$, CQ and RHT tests to the data $\{\widetilde{X}_i\}_{i=1}^{n_1}$ and  $\{\widetilde{Y}_k\}_{k=1}^{n_2}$,
 where BP-DLRT and BP-${\rm GCT}_{\rm lg}$
 are formed by first reordering
 $\widetilde{X}_i$ and $\widetilde{Y}_k$  with the best permutation method, and then applying the DLRT and ${\rm GCT}_{\rm lg}$ to the reordered variables
(note that ${\rm GCT}_{\rm lg}$ also assumes a natural ordering).

The simulation results are shown in Table \ref{tab2} with nominal level of 5\%.
When the correlation is weak, the type I error rates of BP-DLRT are still more close to the nominal level than the BP-${\rm GCT}_{\rm lg}$ test.
The power of the BP-DLRT test is usually larger than
the SD, PA, BP-${\rm GCT}_{\rm lg}$, CQ and RHT tests.
When the correlation becomes stronger, we note that
BP-${\rm GCT}_{\rm lg}$ and BP-${\rm GCT}_{\rm lg}$ both have inflated type I error rates.

In conclusion, when  the natural ordering among the components is unknown,
the BP-DLRT test can be recommended for the un-ordered variables.
In particular, when the correlation among the components are not strong, the BP-DLRT test is able to provide a satisfactory power performance.

\linespread{1}

\begin{center}
{
\begin{table}[htbp!]
\caption{\label{tab2} Power comparisons  over 2000 simulations for the BP-DLRT, BP-${\rm GCT}_{\rm lg}$, PA, SD, CQ and RHT tests under the IND or SRD structure, respectively.  Two
different correlation, $\rho=$ 0.3 or 0.6, are considered for the SRD structure.
The nominal level is 5\%.
}
{{
\begin{center}
\begin{tabular}{c|c||c|c|c||c|c|c}
\hline\hline
& & \multicolumn{3}{c||}{$p=500$} & \multicolumn{3}{c}{$p=800$} \\\hline
\multirow{2}{*}{Cov}& \multirow{2}{*}{Method} & \multirow{2}{*}{$\beta=0$}
& \multirow{2}{*}{$\beta=0.15$}& \multirow{2}{*}{$\beta=0.30$}&
\multirow{2}{*}{$\beta=0$} & \multirow{2}{*}{$\beta=0.15$}& \multirow{2}{*}{$\beta=0.30$}\\
&  & & & & & & \\\hline
\multirow{7}{*}{IND}
&BP-DLRT& 0.051&0.509&0.967
&0.044 &0.695 &0.998
\\\cline{2-8}
%&DLRT&0.051&0.514&0.968& & &   \\\cline{2-8}
&SD&0.032&0.478&0.966
&0.030&0.654  & 0.997 \\\cline{2-8}
&PA& 0.050 &0.544&0.974
&0.045 & 0.745& 0.999 \\\cline{2-8}
&BP-${\rm GCT}_{\rm lg}$ &0.063&0.393&0.947 &0.045 &0.612 &0.995 \\\cline{2-8}
&CQ&0.049&0.466&0.920
&0.051&0.568 &0.979   \\\cline{2-8}
&RHT&0.050&0.434&0.905
& 0.054&0.566 &0.979     \\ \hline
\hline
\multirow{5}{*}{SRD}
&BP-DLRT&0.064&0.536&0.937& 0.071&0.717&0.993 \\\cline{2-8}
&SD&0.039&0.430&0.896&0.032 & 0.585& 0.986  \\\cline{2-8}
&PA&0.048&0.499&0.924&0.061 & 0.678&0.993 \\\cline{2-8}
($\rho=0.3$)
&BP-${\rm GCT}_{\rm lg}$ &0.081& 0.430&0.908 &0.077&0.637 &0.991 \\\cline{2-8}
&CQ&0.049&0.426&0.844& 0.056&0.544 & 0.966 \\\cline{2-8}
&RHT& 0.064 & 0.381& 0.791& 0.058 & 0.502&0.958 \\ \hline\hline
\multirow{5}{*}{SRD}
&BP-DLRT&0.089&0.311&0.722& 0.082&0.486&0.913  \\\cline{2-8}
&SD&0.042&0.251&0.690&0.039 & 0.035& 0.837  \\\cline{2-8}
&PA&0.055&0.343&0.748&0.053 & 0.446&0.901  \\\cline{2-8}
($\rho=0.6$)
&BP-${\rm GCT}_{\rm lg}$ &0.112& 0.265&0.678 &0.093 &0.419 &0.889 \\\cline{2-8}
&CQ&0.054&0.301&0.690& 0.056  &0.359 & 0.836 \\\cline{2-8}
&RHT& 0.056 & 0.235& 0.546& 0.045 & 0.327&0.765 \\ \hline\hline
\end{tabular}\end{center}
} }
\end{table}
}
\end{center}

\linespread{1.25}

\renewcommand\thesection{{D}}{}
\section{Properties of the DLRT, SD, PA, GCT, CQ and RHT tests}

\subsection{A summary of asymptotic power}

{
In this paper we focus on the testing problem with dense but weak signals, the asymptotic power of
the considered tests depend heavily on the total amount of the signals. To specify the necessary order of the respective tests, we have reported their asymptotic power in Table \ref{order}.
In the third column of Table \ref{order}, we present
the necessary order of the signal for each test.}

{
Specifically, we assume $\kappa_0=n_1n_2/(n_1+n_2)$, $\bm{\mu}_1$ and $\bm{\mu}_2$ are two population mean vectors, $\Sigma$ and $R$  be the common covariance matrix and correlation matrix, respectively.
Let $D_{\sigma}={\rm diag}(\sigma_{11}^2,\cdots, \sigma_{pp}^2)$  be the diagonal matrix of $\Sigma$, $\tau_D^2=2\pi f_d(0)$ and $\tau_G^2=2\pi f_g(0)$, where  $f_d(\cdot)$ and  $f_g(\cdot)$
are the spectrum density functions of $\{V_{Nj}\}_{j=1}^p$ and
$\{t_{Nj}\}_{j=1}^p$, respectively.  Let
$$H_{\lambda}=\frac{1}{1+\gamma \Theta_1(\lambda, \gamma)}\Sigma+ \lambda I_p,$$
where
$$\Theta_1(\lambda, \gamma)=\frac{1-\lambda m_F(-\lambda)}{1-\gamma\{1-\lambda m_F(-\lambda)\}},$$
with $\gamma=p/n$,  $\lambda>0$, and $m_F(z)$ is the solution of Marchenko--Pastur equation
\citep{chen2011}.
In addition, let $$\Theta_2(\lambda, \gamma)=\frac{1-\lambda m_F(-\lambda)}{[1-\gamma\{1-\lambda m_F(-\lambda)\}]^3}
-\lambda \frac{m_F(-\lambda)-\lambda m_F'(-\lambda)}{[1-\gamma\{1-\lambda m_F(-\lambda)\}]^4},$$
where $m_F'(z)$ is the derivative of $m_F(z)$.}

{
The asymptotic power and the necessary order of the signal for each test are reported in Table \ref{order}.
}

\begin{table}[htbp!]
\caption{\label{order} Asymptotic power and the necessary order of
signal for the DLRT, SD, PA, GCT, CQ and RHT tests.}
{ {
\begin{center}
\begin{tabular}{c||c|c}
\hline\hline
test method & asymptotic power & the necessary order of signal  \\\hline
\multirow{3}{*}{DLRT}
&\multirow{3}{*}{ $\Phi\Big(-z_{\alpha}+\kappa_0
\frac{(\bm{\mu}_1-\bm{\mu}_2)^TD_{\sigma}^{-1}(\bm{\mu}_1-\bm{\mu}_2)}{\sqrt{p\tau_D^2}}$\Big)}
&\multirow{3}{*}{
${(\bm{\mu}_1-\bm{\mu}_2)^TD_{\sigma}^{-1}(\bm{\mu}_1-\bm{\mu}_2)}=O(\sqrt{p\tau_D^2})$}  \\
& &\\
&  & \\\hline
\multirow{3}{*}{SD} &
\multirow{3}{*}{ $\Phi\Big(-z_{\alpha}+\kappa_0\frac{(\bm{\mu}_1-\bm{\mu}_2)^TD_{\sigma}^{-1}(\bm{\mu}_1-\bm{\mu}_2)}
{\sqrt{2 {\rm tr}{(R^2)}}}$\Big)} &
\multirow{3}{*}{
${(\bm{\mu}_1-\bm{\mu}_2)^TD_{\sigma}^{-1}(\bm{\mu}_1-\bm{\mu}_2)}=O({\sqrt{2 {\rm tr}{(R^2)}}})$} \\
& &\\
  & & \\\hline
\multirow{3}{*}{PA}
& \multirow{3}{*}{ $\Phi\Big(-z_{\alpha}+\kappa_0\frac{(\bm{\mu}_1-\bm{\mu}_2)^T D_{\sigma}^{-1}(\bm{\mu}_1-\bm{\mu}_2)}{\sqrt{2 {\rm tr}{(R^2)}}}$\Big)}
& \multirow{3}{*}{
${(\bm{\mu}_1-\bm{\mu}_2)^TD_{\sigma}^{-1}(\bm{\mu}_1-\bm{\mu}_2)}=O({\sqrt{2 {\rm tr}{(R^2)}}})$} \\
& &\\
&   & \\\hline
\multirow{3}{*}{GCT}
&\multirow{3}{*}{
$\Phi\Big( -z_{\alpha}+\kappa_0\frac{(\bm{\mu}_1-\bm{\mu}_2)^T
D_{\sigma}^{-1} (\bm{\mu}_1-\bm{\mu}_2)}{\sqrt{p\tau_G^2}} \Big)$}
& \multirow{3}{*}{
${(\bm{\mu}_1-\bm{\mu}_2)^TD_{\sigma}^{-1}(\bm{\mu}_1-\bm{\mu}_2)}=O({\sqrt{p\tau_G^2}}) $}  \\
& &\\
&  &\\\hline
\multirow{3}{*}{CQ}
&\multirow{3}{*}{ $\Phi\Big(-z_{\alpha}+\kappa_0\frac{(\bm{\mu}_1-\bm{\mu}_2)^TD_{\sigma}^{-1}(\bm{\mu}_1-\bm{\mu}_2)}{\sqrt{2 {\rm tr}{(\Sigma^2)}}}$ \Big)}
&\multirow{3}{*}{
${(\bm{\mu}_1-\bm{\mu}_2)^TD_{\sigma}^{-1}(\bm{\mu}_1-\bm{\mu}_2)}=O({\sqrt{2 {\rm tr}{(\Sigma^2)}}})$} \\
& &\\
  & & \\\hline
\multirow{3}{*}{RHT} &
\multirow{3}{*}{
$\Phi\Big(-z_{\alpha}+\kappa_0\frac{(\bm{\mu}_1-\bm{\mu}_2)^TH_{\lambda}^{-1}
(\bm{\mu}_1-\bm{\mu}_2)}{\sqrt{2p \Theta_2(\lambda,\gamma)}}$\Big)}
&\multirow{3}{*}{
${(\bm{\mu}_1-\bm{\mu}_2)^TH_{\lambda}^{-1}
(\bm{\mu}_1-\bm{\mu}_2)}=O({\sqrt{2p \Theta_2(\lambda,\gamma)}})$} \\
& &\\
& & \\\hline\hline
\end{tabular}\end{center}
} }
\end{table}

\subsection{A summary of the transformation invariance properties}
{
The transformation invariant is a desirable property when constructing the test statistics. To cater for demand, let $\bm{X} \in R^p$ is a random vector, we consider the following three types of transformation:
\begin{itemize}
  \item[(1)] the orthogonal transformation: $\bm{X} \to \Gamma \bm{X}$, where $\Gamma^T\Gamma=I $ is an identity matrix;
  \item[(2)] the scale transformation: $\bm{X} \to D \bm{X}$, where $D={\rm diag}(d_1,\ldots,d_p)$ without any zero entries;
 \item[(3)] the shift transformation: $\bm{X} \to \bm{d}_0+ \bm{X}$, where $\bm{d}_0\in R^p$ is a constant vector.
\end{itemize}
}

{
We summarize the transformation invariance properties of the DLRT, SD, PA, GCT, CQ and RHT tests into Table \ref{tab1}. Interestingly,  none of the six tests possesses all the three invariance properties.
Future research may be warranted to investigate
whether there exists a new test that possesses all the three invariance properties.}

%How to construct test statistics satisfied more invariance properties deserves future study.

\begin{table}[htbp!]
\caption{\label{tab1}The transformation invariance properties of the DLRT, SD, PA, GCT, CQ and RHT tests}
{{
\begin{center}
\begin{tabular}{c|c|c|c|c|c|c}
\hline\hline
 & DLRT &SD& PA& GCT& CQ& RHT  \\\hline
orthogonal invariance&No &No&No&No& Yes &Yes \\\hline
scale invariance& Yes& Yes & Yes& Yes&No & No \\ \hline
shift invariance& Yes& Yes & No& Yes&No & Yes \\ \hline \hline
\end{tabular}\end{center}
} }
\end{table}

\clearpage
\bibliographystyle{biom}
\bibliography{LRT-reference}

\begin{thebibliography}{}

\bibitem[\protect\citeauthoryear{Ahmad}{Ahmad}{2014}]{ahmad2014}
Ahmad, M.~R. (2014).
\newblock A {$U$}-statistic approach for a high-dimensional two-sample mean
  testing problem under non-normality and {B}ehrens-{F}isher setting.
\newblock {\em The Annals of the Institute of Statistical Mathematics} {\bf
  66,} 33--61.

\bibitem[\protect\citeauthoryear{Bai and Saranadasa}{Bai and
  Saranadasa}{1996}]{bai1996}
Bai, Z. and Saranadasa, H. (1996).
\newblock Effect of high dimension: by an example of a two sample problem.
\newblock {\em Statistica Sinica} {\bf 6,} 311--329.

\bibitem[\protect\citeauthoryear{Baladandayuthapani, Ji, Talluri,
  Nieto-Barajas, and Morris}{Baladandayuthapani
  et~al.}{2010}]{Baladandayuthapani2010}
Baladandayuthapani, V., Ji, Y., Talluri, R., Nieto-Barajas, L.~E., and Morris,
  J.~S. (2010).
\newblock {Bayesian random segmentation models to identify shared copy number
  aberrations for array {CGH} data}.
\newblock {\em Journal of the American Statistical Association} {\bf 105,}
  1358--1375.

\bibitem[\protect\citeauthoryear{Bickel and Levina}{Bickel and
  Levina}{2004}]{bickel2004}
Bickel, P.~J. and Levina, E. (2004).
\newblock Some theory for {F}isher's linear discriminant function, `naive
  {B}ayes', and some alternatives when there are many more variables than
  observations.
\newblock {\em Bernoulli} {\bf 10,} 989--1010.

\bibitem[\protect\citeauthoryear{B{\"u}hlmann}{B{\"u}hlmann}{1996}]{buhlmann1996}
B{\"u}hlmann, P. (1996).
\newblock Locally adaptive lag--window spectral estimation.
\newblock {\em Journal of Time Series Analysis} {\bf 17,} 247--270.

\bibitem[\protect\citeauthoryear{Chakraborty and Chaudhuri}{Chakraborty and
  Chaudhuri}{2017}]{chakraborty2017}
Chakraborty, A. and Chaudhuri, P. (2017).
\newblock Tests for high-dimensional data based on means, spatial signs and
  spatial ranks.
\newblock {\em The Annals of Statistics} {\bf 45,} 771--799.

\bibitem[\protect\citeauthoryear{Chen, Paul, Prentice, and Wang}{Chen
  et~al.}{2011}]{chen2011}
Chen, L.~S., Paul, D., Prentice, R.~L., and Wang, P. (2011).
\newblock A regularized {H}otelling's {$T^2$} test for pathway analysis in
  proteomic studies.
\newblock {\em Journal of the American Statistical Association} {\bf 106,}
  1345--1360.

\bibitem[\protect\citeauthoryear{Chen and Qin}{Chen and Qin}{2010}]{chen2010}
Chen, S. and Qin, Y. (2010).
\newblock A two-sample test for high-dimensional data with applications to
  gene-set testing.
\newblock {\em The Annals of Statistics} {\bf 38,} 808--835.

\bibitem[\protect\citeauthoryear{Dong, Pang, Tong, and Genton}{Dong
  et~al.}{2016}]{Dong2015}
Dong, K., Pang, H., Tong, T., and Genton, M.~G. (2016).
\newblock Shrinkage-based diagonal {H}otelling's tests for high-dimensional
  small sample size data.
\newblock {\em Journal of Multivariate Analysis} {\bf 143,} 127--142.

\bibitem[\protect\citeauthoryear{Dudoit, Fridlyand, and Speed}{Dudoit
  et~al.}{2002}]{dudoit2002}
Dudoit, S., Fridlyand, J., and Speed, T.~P. (2002).
\newblock Comparison of discrimination methods for the classification of tumors
  using gene expression data.
\newblock {\em Journal of the American Statistical Association} {\bf 97,}
  77--87.

\bibitem[\protect\citeauthoryear{Feng, Zou, Wang, and Zhu}{Feng
  et~al.}{2015}]{feng2015}
Feng, L., Zou, C., Wang, Z., and Zhu, L. (2015).
\newblock Two-sample {B}ehrens-{F}isher problem for high-dimensional data.
\newblock {\em Statistica Sinica} {\bf 25,} 1297--1312.

\bibitem[\protect\citeauthoryear{Ghosh and Biswas}{Ghosh and
  Biswas}{2016}]{ghosh2016}
Ghosh, A.~K. and Biswas, M. (2016).
\newblock Distribution-free high-dimensional two-sample tests based on
  discriminating hyperplanes.
\newblock {\em TEST} {\bf 25,} 525--547.

\bibitem[\protect\citeauthoryear{Gilbert, Moler, and Schreiber}{Gilbert
  et~al.}{1992}]{gilbert1992}
Gilbert, J.~R., Moler, C., and Schreiber, R. (1992).
\newblock Sparse matrices in {MATLAB}: design and implementation.
\newblock {\em SIAM Journal on Matrix Analysis and Applications} {\bf 13,}
  333--356.

\bibitem[\protect\citeauthoryear{Gregory, Carroll, Baladandayuthapani, and
  Lahiri}{Gregory et~al.}{2015}]{Carroll2015}
Gregory, K.~B., Carroll, R.~J., Baladandayuthapani, V., and Lahiri, S.~N.
  (2015).
\newblock A two-sample test for equality of means in high dimension.
\newblock {\em Journal of the American Statistical Association} {\bf 110,}
  837--849.

\bibitem[\protect\citeauthoryear{Hall and Heyde}{Hall and
  Heyde}{1980}]{hall2014}
Hall, P. and Heyde, C.~C. (1980).
\newblock {\em Martingale {L}imit {T}heory and {I}ts {A}pplication}.
\newblock Academic Press, New York.

\bibitem[\protect\citeauthoryear{Huang, Tong, and Zhao}{Huang
  et~al.}{2010}]{huang2010}
Huang, S., Tong, T., and Zhao, H. (2010).
\newblock Bias-corrected diagonal discriminant rules for high-dimensional
  classification.
\newblock {\em Biometrics} {\bf 66,} 1096--1106.

\bibitem[\protect\citeauthoryear{Jiang and Qi}{Jiang and Qi}{2015}]{Jiang2015}
Jiang, T. and Qi, Y. (2015).
\newblock Likelihood ratio tests for high-himensional normal distributions.
\newblock {\em Scandinavian Journal of Statistics} {\bf 42,} 988--1009.

\bibitem[\protect\citeauthoryear{Jiang and Yang}{Jiang and
  Yang}{2013}]{Jiang2013}
Jiang, T. and Yang, F. (2013).
\newblock Central limit theorems for classical likelihood ratio tests for
  high-dimensional normal distributions.
\newblock {\em The Annals of Statistics} {\bf 41,} 2029--2074.

\bibitem[\protect\citeauthoryear{Li, Aue, Paul, Peng, and Wang}{Li
  et~al.}{2016}]{li2016}
Li, H., Aue, A., Paul, D., Peng, J., and Wang, P. (2016).
\newblock An adaptable generalization of {H}otelling's ${T}^{2}$ test in high
  dimension.
\newblock {\em arXiv preprint arXiv:1609.08725} .

\bibitem[\protect\citeauthoryear{Lopes, Jacob, and Wainwright}{Lopes
  et~al.}{2011}]{lopes2011}
Lopes, M.~E., Jacob, L., and Wainwright, M.~J. (2011).
\newblock A more powerful two-sample test in high dimensions using random
  projection.
\newblock {\em Advances in Neural Information Processing Systems} {\bf 24,}
  1206--1214.

\bibitem[\protect\citeauthoryear{Olshen, Venkatraman, Lucito, and
  Wigler}{Olshen et~al.}{2004}]{olshen2004}
Olshen, A.~B., Venkatraman, E., Lucito, R., and Wigler, M. (2004).
\newblock Circular binary segmentation for the analysis of array-based {DNA}
  copy number data.
\newblock {\em Biostatistics} {\bf 5,} 557--572.

\bibitem[\protect\citeauthoryear{Paparoditis and Politis}{Paparoditis and
  Politis}{2012}]{paparoditis2012}
Paparoditis, E. and Politis, D.~N. (2012).
\newblock Nonlinear spectral density estimation: thresholding the correlogram.
\newblock {\em Journal of Time Series Analysis} {\bf 33,} 386--397.

\bibitem[\protect\citeauthoryear{Park and Ayyala}{Park and
  Ayyala}{2013}]{park2013test}
Park, J. and Ayyala, D.~N. (2013).
\newblock A test for the mean vector in large dimension and small samples.
\newblock {\em Journal of Statistical Planning and Inference} {\bf 143,}
  929--943.

\bibitem[\protect\citeauthoryear{Parzen}{Parzen}{1961}]{Parzen1961}
Parzen, E. (1961).
\newblock Mathematical considerations in the estimation of spectra.
\newblock {\em Technometrics} {\bf 3,} 167--190.

\bibitem[\protect\citeauthoryear{Rajaratnam and Salzman}{Rajaratnam and
  Salzman}{2013}]{rajaratnam2013}
Rajaratnam, B. and Salzman, J. (2013).
\newblock Best permutation analysis.
\newblock {\em Journal of Multivariate Analysis} {\bf 121,} 193--223.

\bibitem[\protect\citeauthoryear{Spira}{Spira}{1971}]{spira1971}
Spira, R. (1971).
\newblock Calculation of the gamma function by {S}tirling's formula.
\newblock {\em Mathematics of Computation} {\bf 25,} 317--322.

\bibitem[\protect\citeauthoryear{Srivastava}{Srivastava}{2009}]{Srivastava2009}
Srivastava, M.~S. (2009).
\newblock A test for the mean vector with fewer observations than the dimension
  under non-normality.
\newblock {\em Journal of Multivariate Analysis} {\bf 100,} 518--532.

\bibitem[\protect\citeauthoryear{Srivastava and Du}{Srivastava and
  Du}{2008}]{srivastava2008}
Srivastava, M.~S. and Du, M. (2008).
\newblock A test for the mean vector with fewer observations than the
  dimension.
\newblock {\em Journal of Multivariate Analysis} {\bf 99,} 386--402.

\bibitem[\protect\citeauthoryear{Srivastava, Katayama, and Kano}{Srivastava
  et~al.}{2013}]{srivastava2013}
Srivastava, M.~S., Katayama, S., and Kano, Y. (2013).
\newblock A two sample test in high dimensional data.
\newblock {\em Journal of Multivariate Analysis} {\bf 114,} 349--358.

\bibitem[\protect\citeauthoryear{Srivastava, Li, and Ruppert}{Srivastava
  et~al.}{2016}]{srivastava2015}
Srivastava, R., Li, P., and Ruppert, D. (2016).
\newblock {RAPTT}: an exact two-sample test in high dimensions using random
  projections.
\newblock {\em Journal of Computational and Graphical Statistics} {\bf 23,}
  954--970.

\bibitem[\protect\citeauthoryear{St{\"a}dler and Mukherjee}{St{\"a}dler and
  Mukherjee}{2017}]{stadler2016}
St{\"a}dler, N. and Mukherjee, S. (2017).
\newblock Two-sample testing in high dimensions.
\newblock {\em Journal of the Royal Statistical Society: Series B} {\bf 79,}
  225--246.

\bibitem[\protect\citeauthoryear{Thulin}{Thulin}{2014}]{thulin2014}
Thulin, M. (2014).
\newblock A high-dimensional two-sample test for the mean using random
  subspaces.
\newblock {\em Computational Statistics \& Data Analysis} {\bf 74,} 26--38.

\bibitem[\protect\citeauthoryear{Wagaman and Levina}{Wagaman and
  Levina}{2009}]{wagaman2009}
Wagaman, A. and Levina, E. (2009).
\newblock Discovering sparse covariance structures with the {I}somap.
\newblock {\em Journal of Computational and Graphical Statistics} {\bf 18,}
  551--572.

\bibitem[\protect\citeauthoryear{Wang, Peng, and Li}{Wang
  et~al.}{2015}]{wang2015}
Wang, L., Peng, B., and Li, R. (2015).
\newblock A high-dimensional nonparametric multivariate test for mean vector.
\newblock {\em Journal of the American Statistical Association} {\bf 110,}
  1658--1669.

\bibitem[\protect\citeauthoryear{Wei, Lee, Wichers, and Marron}{Wei
  et~al.}{2016}]{wei2016}
Wei, S., Lee, C., Wichers, L., and Marron, J. (2016).
\newblock Direction-projection-permutation for high-dimensional hypothesis
  tests.
\newblock {\em Journal of Computational and Graphical Statistics} {\bf 25,}
  549--569.

\bibitem[\protect\citeauthoryear{Wu, Genton, and Stefanski}{Wu
  et~al.}{2006}]{Wu2006}
Wu, Y., Genton, M.~G., and Stefanski, L.~A. (2006).
\newblock A multivariate two-sample mean test for small sample size and missing
  data.
\newblock {\em Biometrics} {\bf 62,} 877--885.

\bibitem[\protect\citeauthoryear{Zhao and Xu}{Zhao and Xu}{2016}]{zhao2016}
Zhao, J. and Xu, X. (2016).
\newblock A generalized likelihood ratio test for normal mean when $p$ is
  greater than $n$.
\newblock {\em Computational Statistics \& Data Analysis} {\bf 99,} 91--104.

\bibitem[\protect\citeauthoryear{Zhu and Galbraith}{Zhu and
  Galbraith}{2010}]{zhu2010}
Zhu, D. and Galbraith, J.~W. (2010).
\newblock A generalized asymmetric {S}tudent-$t$ distribution with application
  to financial econometrics.
\newblock {\em Journal of Econometrics} {\bf 157,} 297--305.

\end{thebibliography}
\clearpage

\end{document}